\documentclass[prb,twocolumn,showpacs]{revtex4}
\usepackage[latin1]{inputenc}
\usepackage{graphicx}
\usepackage{amssymb}
\usepackage{amsmath}
\usepackage{xspace}
\usepackage{dcolumn}
\usepackage{bm}
\usepackage{citesort}

\newcommand{\ie}[0]{i.e.\@\xspace}
\newcommand{\eg}[0]{e.g.\@\xspace}

\newcommand{\Z}[0]{\mathcal{Z}}
\newcommand{\D}[0]{\mathcal{D}}
\newcommand{\U}[0]{\mathcal{U}}
\newcommand{\V}[0]{\mathcal{V}}

\newcommand{\rD}[0]{\text{D}}

\newcommand{\etal}[0]{et al.\@\xspace}
\newcommand{\tr}[0]{\text{Tr}\,}
\newcommand{\om}[0]{\omega}
\newcommand{\si}[0]{\sigma}
\newcommand{\las}[0]{\langle}
\newcommand{\ras}[0]{\rangle}
\newcommand{\la}[0]{\left\las}
\newcommand{\ra}[0]{\right\ras}
\newcommand{\ket}[1]{\left|#1\ra}  
\newcommand{\bra}[1]{\la#1\right|} 
\newcommand{\dtau}{\Delta\tau}
\newcommand{\rmd}{\mathrm{d}}
\newcommand{\rmi}{\mathrm{i}}

\newcommand{\op}{\hat{p}}
\newcommand{\ox}{\hat{x}}
\newcommand{\on}{\hat{n}}
\newcommand{\wb}{w_\text{b}}
\newcommand{\wf}{w_\text{f}}
\newcommand{\UP}{\uparrow}
\newcommand{\DO}{\downarrow}

\newcommand{\Ep}{E_\text{P}}
\newcommand{\Ek}{E_\text{k}}
\newcommand{\Ekb}{\overline{E}_\text{k}}
\newcommand{\nag}{\phantom{\dag}}

\newcommand{\omb}[0]{\overline{\omega}}
\newcommand{\Ub}[0]{\overline{U}}

\DeclareMathAlphabet{\bi}{OML}{cmm}{b}{it}
\newfont{\tensy}{cmsy10}

\begin{document}


\title{Temperature- and quantum phonon effects on Holstein-Hubbard bipolarons}

\author{Martin Hohenadler} \email{hohenadler@itp.tu-graz.ac.at}
\author{Wolfgang \surname{von der Linden}}
\affiliation{%
  Institute for Theoretical and Computational Physics, Graz University of
  Technology, Petersgasse 16, 8010 Graz, Austria}

\begin{abstract}
  The one-dimensional Holstein-Hubbard model with two electrons of opposite
  spin is studied using an extension of a recently developed quantum Monte
  Carlo method, and a very simple yet rewarding variational approach, both
  based on a canonically transformed Hamiltonian.  The quantum Monte Carlo
  method yields very accurate results in the regime of small but finite
  phonon frequencies, characteristic of many strongly correlated
  materials such as, e.g., the cuprates and the manganites. The influence of
  electron-electron repulsion, phonon frequency and temperature on the
  bipolaron state is investigated. Thermal dissociation of the intersite
  bipolaron is observed at high temperatures, and its relation to an existing
  theory of the manganites is discussed.
\end{abstract}

\pacs{63.20.Kr, 71.27.+a, 71.38.-k, 71.38.Mx 02.70.Ss}

\maketitle

%
%
%
\section{\label{sec:introduction}Introduction}
%
%
%

In recent years, the formation and properties of bipolarons, consisting of
two electrons forming a pair in real space, have received considerable
interest due to their potential role, \eg, in high-temperature
superconductivity. Theories based on bipolaron formation have been proposed
for the superconducting transition in the cuprates,\cite{AlKaMo96} and the
metal-insulator transition and colossal magnetoresistance in the
manganites.\cite{AlBr99prl,AlBr99} Despite some fundamental
problems,\cite{ChRaFe98,ChRaFe99,David_AiP} they are still issue of ongoing
discussion.

Many interesting materials fall into the adiabatic regime of small but finite
phonon frequencies and intermediate to strong electron-phonon coupling. For such
parameters, analytical approaches based on, \eg, perturbation theory, do not
give reliable results. In contrast, computational methods represent a very
powerful instrument to obtain exact, unbiased information, and a lot of
numerical work has recently been devoted to an understanding of the Holstein
and the Holstein-Hubbard (HH) model.

In this paper, we present a simple but surprisingly accurate variational
approach to the HH bipolaron. More importantly, we extend a recently
developed quantum Monte Carlo (QMC) method\cite{HoEvvdL03} to the case of two
electrons of opposite spin. The resulting algorithm is used to study
bipolaron formation in the one-dimensional HH model, focusing on the
adiabatic regime. While the ground-state properties of the HH bipolaron are
rather well understood, here we exploit the capability of the QMC approach to also
study finite temperatures. We find that, in particular, the weakly bound
intersite bipolaron is susceptible to thermal dissociation. Furthermore, in
contrast to previous studies, we are able to consider a very large range of
the electron-phonon and electron-electron interaction.

The outline of this work is as follows. In Sec.~\ref{sec:holstein} we discuss
the HH model with two electrons, while in Sec.~\ref{sec:LFT} we present an
extended Lang-Firsov transformation with nonlocal lattice displacements.
Section~\ref{sec:QMC} features the extension of the QMC method to the
bipolaron problem, and Sec.~\ref{sec:VPA} covers the variational approach.
Results are presented in Sec.~\ref{sec:results}, and Sec.~\ref{sec:summary}
contains our conclusions.

%
%
%
\section{\label{sec:holstein}The Holstein-Hubbard model}
%
%
%

The HH model is defined in terms of dimensionless phonon
by the Hamiltonian
\begin{eqnarray}\label{eq:holstein}\nonumber
  H
  &=&
  \underbrace{%
  -t\sum_{\las ij\ras\si} c^\dag_{i\si} c^{\nag }_{j\si}}
  _{K}
  +
  \underbrace{%
  \frac{\om}{2}\sum_i\left( \op_i^2 + \ox_i^2 \right)}
  _{P=P_p + P_x} 
  \\
  &&
  \underbrace{%
  -\alpha\sum_i \on_i \ox_i}
  _{I_\text{ep}}
  +
  \underbrace{%
  U \sum_i \on_{i\UP} \on_{i\DO}}
  _{I_\text{ee}}
  \,,
\end{eqnarray}
where $K$ describes the hopping of electrons, $P$ corresponds to the sum of
the kinetic ($P_p$) and elastic ($P_x$) energy of the phonons, and
$I_\text{ep}$, $I_\text{ee}$ denote the electron-phonon (el-ph) and
electron-electron (el-el) interaction terms, respectively. Here
$c^\dag_{i\sigma}$ ($c_{i\sigma}$) creates (annihilates) an electron of spin
$\si$ at lattice site $i$, $\ox_i$ and $\op_i$ denote the displacement and
momentum of a harmonic oscillator at site $i$, and $\on_i=\sum_\si
\on_{i\si}$ with $\on_{i\si}=c^\dag_{i\si} c^{\nag }_{i\si}$. The third term,
$I_\text{ep}$, describes the coupling of dispersionless Einstein phonons to
the electron occupation number $\on_i$. For doped cuprates or manganites,
such a local interaction is expected to be a reasonable approximation as a
result of screening. In the first term, the symbol $\las ij\ras$ denotes a
summation over all nearest-neighbor hopping pairs $(i,j)$ and $(j,i)$.  The
parameters of the model are the hopping integral $t$, the phonon energy $\om$
($\hbar=1)$, the el-ph coupling constant $\alpha$, and the Coulomb repulsion
$U>0$.  For $U=0$, Eq.~(\ref{eq:holstein}) is identical to the Holstein
model.\cite{Ho59a} As in previous work, we introduce the dimensionless
coupling constant $\lambda =\alpha^2/(\om W)$, where $W=4t\rD$ is the bare
bandwidth in D dimensions. We further define the parameters $\omb=\om/t$ and
$\Ub=U/t$, and express all energies in units of $t$. Consequently, the
independent parameters of the model are $\omb$, $\lambda$, and $\Ub$. We
shall see below that a very useful quantity is given by the polaron binding
energy $\Ep=\lambda W /2$.  Finally, throughout this paper, periodic boundary
conditions in real space are assumed.

This work is exclusively concerned with the case of two electrons, neglecting
the interaction between bipolarons which will definitely be present to some
degree in real materials.  Furthermore, we restrict our attention to two
electrons with opposite spin, \ie, to the singlet bipolaron. A comparison of
the singlet and triplet state has recently been given in
Ref.~\onlinecite{HoAivdL04}.

A review of early work on the bipolaron problem can be found in the book of
Alexandrov and Mott.\cite{AlMo95} Here we focus the discussion on more recent
developments. The latter can roughly be divided into two classes depending on
the methods employed: (a) variational
approaches,\cite{CoEcSo84,EmYeBe92,LaMaPu97,PrAu98,PrAu99,PrAu00,deFiCaIaMaPeVe01}
and (b) unbiased numerical studies using
ED,\cite{RaTh92,Marsiglio95,WeRoFe96,dMeRa97,dMeRa98} variational
diagonalization,\cite{FeRoWeMi95,WeFeWeBi00,BoKaTr00,ElShBoKuTr03}
the density-matrix renormalization group (DMRG),\cite{ZhJeWh99} and
QMC.\cite{deRaLa86,Mac04} Except for the QMC study
of de Raedt and Lagendijk,\cite{deRaLa86} all work was restricted to the
ground state. Moreover, even their QMC results were reported only for a
single, low temperature. This motivates our study of temperature effects
in Sec.~\ref{sec:QMC}.

While ED and DMRG studies were obtained on clusters with
two,\cite{RaTh92,dMeRa97,dMeRa98} four,\cite{Marsiglio95} six,\cite{ZhJeWh99}
eight\cite{FeRoWeMi95,WeRoFe96} or twelve sites,\cite{WeFeWeBi00} the
variational methods of Bon\v{c}a \etal\cite{BoKaTr00} and of
Refs.~\onlinecite{CoEcSo84,EmYeBe92,LaMaPu97,PrAu98,PrAu99,PrAu00,deFiCaIaMaPeVe01}
are only weakly influenced by finite-size effects.  An important disadvantage
of ED and DMRG is the fact that the phonon Hilbert space has to be truncated,
so that these methods can not easily be used to study the adiabatic
($\omb\ll1$) and/or strong-coupling ($\lambda\gg1$) regime. In contrast, no
such limitations are imposed on QMC and most variational methods.

Although de Raedt and Lagendijk only considered the adiabatic limit $\omb=0$,
similar to other authors,\cite{CoEcSo84,deRaLa86,EmYeBe92,PrAu98,PrAu99}
their method can also be applied for finite phonon frequency.\cite{deRaLa86}
Moreover, it may be generalized to include dispersive phonons. Recently, an
extended Holstein model with long-range el-ph interaction has been
investigated by Bon\v{c}a and Trugman.\cite{BoTr01} De Raedt and Lagendijk
also considered long-range Coulomb interaction, while most other authors only
took into account the local Hubbard-type interaction given in
Eq.~(\ref{eq:holstein}), except for Zhang \etal\cite{ZhJeWh99} who have
omitted this term in their DMRG calculations. Finally, we would like to point
out that bipolaron formation in a model with Jahn-Teller modes---as present
in the perovskite manganites---has been studied by Shawish
\etal\cite{ElShBoKuTr03} 

%
%
%
\section{\label{sec:LFT} Transformed Hamiltonians}
%
%
%

The basis of both the variational approach and the QMC method presented below
is the unitary transformation $\tilde{H}\equiv\nu H \nu^\dag$ of the
Hamiltonian~(\ref{eq:holstein}), with
$\nu=\exp(\rmi\sum_{ij}\gamma_{ij}\on_i\op_j)$ (see
Ref.~\onlinecite{HoEvvdL03}). The result is
\begin{eqnarray}\label{eq:htilde}
  \tilde{H}
  &=&
  \underbrace{%
  -t\sum_{\las ij\ras\si}
  c^\dag_{i\si}c^{\nag }_{j\si}
  e^{\rmi\sum_l(\gamma_{il}-\gamma_{jl})\op_l}}
  _{\tilde{K}}
  + 
  P
  \\\nonumber
  &&+
  \underbrace{%
  \sum_{ij} \on_j
  \ox_i(\om\gamma_{ij}-\alpha\delta_{ij})}
  _{\tilde{I}_\text{ep}}
  +
  \underbrace{%
  \sum_{ij} v_{ij} \on_i \on_j - \frac{U}{2}\sum_i \on_i}
  _{\tilde{I}_\text{ee}}
  \,,
\end{eqnarray}
with
\begin{equation}\label{eq:htilde_el-el}
  v_{ij}
  =
  \frac{\omega}{2}
  \sum_l\gamma_{lj}\gamma_{li}-\alpha\gamma_{ij}
  + \frac{1}{2}\delta_{ij} U
  \,.
\end{equation}
As discussed in Ref.~\onlinecite{HoEvvdL03}, henceforth also referred to as
I, the extended transformation $\nu$ takes into account nonlocal
lattice displacements, which are essential for a correct description in the
regime $\omb\lesssim1$.

Similar to I, for the QMC method, we resort to the
standard Lang-Firsov (LF) transformation\cite{LangFirsov} with
$\nu_0=\exp(\rmi\gamma\sum_i\on_i\op_i)$. Here $\gamma=\sqrt{\lambda W/\om}$
has been chosen such that the el-ph coupling term $I_\text{ep}$ in
Eq.~(\ref{eq:holstein}) cancels. The transformed Hamiltonian then takes the form
\begin{eqnarray}\label{eq:LFHamiltonian}\nonumber
  \tilde{H}_0
  &=&
  \underbrace{%
  -t\sum_{\las ij\ras\si}c^\dag_{i\si}c^{\nag }_{j\si}
  e^{\rmi\gamma(\op_i-\op_j)}}
  _{\tilde{K}_0}
   + P
  \\
  &&+
  \underbrace{%
  (U-2\Ep )\sum_i \on_{i\UP}\on_{i\DO}}
  _{\tilde{I}}
  - 2\Ep
  \,.
\end{eqnarray}
Hence, in contrast to the polaron problem,\cite{HoEvvdL03} the el-el
interaction term, resulting from the canonical transformation, does not
vanish but instead combines with the Hubbard term.

%
%
%
\section{\label{sec:QMC} Quantum Monte Carlo}
%
%
%

The derivation of the QMC algorithm for the bipolaron problem is very similar
to the one-electron case,\cite{HoEvvdL03} and we shall therefore focus on the
differences occurring. Moreover, we restrict the discussion to one dimension.

%
%
\subsection{Partition function}\label{sec:partition-function}
%
%

We set out to calculate the partition function $\Z=e^{-\beta\tilde{H}_0}$, with
$\tilde{H}_0$ given by Eq.~(\ref{eq:LFHamiltonian}). To this end, we first
notice that the last term in Hamiltonian~(\ref{eq:LFHamiltonian}) is a constant and
can therefore be neglected during the QMC simulation. Using the standard
Suzuki-Trotter decomposition we obtain\cite{wvl1992}
\begin{equation}\label{eq:suzuki-trotter}
  e^{-\beta\tilde{H}_0}
  \approx
  \left(
    e^{-\dtau\tilde{K}_0}
    e^{-\dtau P_\text{p}}
    e^{-\dtau P_\text{x}}
    e^{-\dtau \tilde{I}}
  \right)^L \equiv \U^L\,,
\end{equation}
where $\beta=(k_\text{B}T)^{-1}$ and $\dtau=\beta/L$. Inserting $L$ complete
sets of phonon momentum eigenstates and splitting up the trace into a bosonic and a
fermionic part we find\cite{HoEvvdL03}
\begin{equation}
  \Z_L
  =
  \tr_\text{f}\int\,\rmd p_1\rmd p_2\cdots\rmd p_L
  \bra{p_1}\U\ket{p_2}\cdots\bra{p_L}\U\ket{p_1}\;,
\end{equation}
where $\rmd p_\tau\equiv\prod_i \rmd p_{i,\tau}$, and
$\lim_{L\rightarrow\infty}\Z_L=\Z$.\cite{wvl1992} Since the phonon
contribution to $\U$ is identical to the single-electron problem,\cite{HoEvvdL03}
we can again integrate out the coordinates $\ox$. Upon defining $\D p=\rmd
p_1\rmd p_2\cdots\rmd p_L$ the partition function becomes
\begin{equation}\label{eq:Z}
  \Z_L
   =
   C \int\,\D p\,\, \wb \,\wf,
\end{equation}
with $C=[2\pi/(\om\dtau)]^{NL}$,
\begin{eqnarray}\label{eq:omega}\nonumber
   \wb
   &=&
   e^{-\dtau S_\text{b}}
   \,,\quad
   \wf
   =
   \tr_\text{f}\,\Omega
   \\
   \Omega
   &=&
   \prod_{\tau=1}^L
   e^{-\dtau\tilde{K}_{0,\tau}}e^{-\dtau\tilde{I}}
   \,.
\end{eqnarray}
Here $\tilde{K}_{0,\tau}$ is obtained from $\tilde{K}_0$ [Eq.~(\ref{eq:LFHamiltonian})] by replacing $\op_i$
($\op_j$) with $p_{i,\tau}$ ($p_{j,\tau}$). The bosonic action  has the form 
\begin{equation}\label{eq:action-w-matrix}
  S_\text{b}
  =
  \sum_{i=1}^N \bm{p}_i^\text{T} A \,\bm{p}_i
  \,,
\end{equation}
with $\bm{p}_i=(p_{i,1},\dots p_{i,L})$ and a tridiagonal $L\times L$ matrix
$A$ defined by
\begin{equation}\label{eq:matrixA}
  A_{l,l}
  =\frac{\om}{2}+\frac{1}{\om\dtau^2}\;,\quad
  A_{l,l\pm1}
  =
  -\frac{1}{2\om\dtau^2}\,.
\end{equation}
As pointed out in I, the representation of $S_\text{b}$ given in
Eq.~(\ref{eq:action-w-matrix}) permits us to introduce the so-called principal
component representation discussed below.

To evaluate the fermionic trace  we choose the
two-electron basis states
\begin{equation}\label{eq:basis}
  \left\{
    \ket{l}
    \equiv
    \ket{i,j}
    \equiv
    c^\dag_{i\UP}c^\dag_{j\DO}\ket{0}
    \,,\quad
    i,j = 1,\dots,N
  \right\}
  \,,
\end{equation}
where we have introduced a combined index $l$ running from 1 to $N^2$ in one
dimension. We begin with the contribution of the kinetic term $\tilde{K}_0$
[Eq.~(\ref{eq:LFHamiltonian})].  It follows that the tight-binding hopping
matrix, denoted as $\kappa$, has dimension $N^2\times N^2$.  The exponential
of the transformed hopping term can be written as\cite{HoEvvdL03}
\begin{equation}
  e^{-\dtau\tilde{K}_{0,\tau}}
  =
  D_\tau \kappa D_\tau^\dag
  \,,
\end{equation}
where
\begin{equation}
  (D_\tau)_{ll'}
  =
  \delta_{ll'}
  (\delta_{n_{i\UP},1}\delta_{n_{j\DO},1}+\delta_{n_{i\DO},1}\delta_{n_{j\UP},1})
  e^{\rmi\gamma(p_{i,\tau}+p_{j,\tau})}
\end{equation}
is diagonal in the basis~(\ref{eq:basis}).

The second contribution to the matrix $\Omega$ in Eq.~(\ref{eq:omega}) comes
from the effective el-el interaction term $\tilde{I}$
[Eq.~(\ref{eq:LFHamiltonian})] in terms of the diagonal matrix
\begin{equation}
  (\V_\tau)_{ll'}
  =
  \delta_{ll'}\;e^{\dtau(U-2\Ep )\delta_{ij}}
  \,.
\end{equation}
We would like to emphasize that
the random variables $\bm{p}$ merely enter the diagonal matrix $D$, 
while the $N^2\times N^2$ matrices $\V_\tau$ and $\kappa$ are fixed
throughout the entire MC simulation. Thus, in total, we have
\begin{equation}\label{eq:matproduct}
  \Omega
  =
  \prod_\tau D_\tau \kappa D^\dag_\tau \V_\tau
  \,,
\end{equation}
and the fermionic trace is calculated as
\begin{equation}
  \tr_\text{f}\,\Omega
  =
  \sum_{ij} \bra{i,j} \Omega \ket{i,j}
  \,,
\end{equation}
which is identical to the sum over the diagonal elements of the matrix
$\Omega$ in the basis~(\ref{eq:basis}).

%
%
\subsection{Observables}\label{sec:observables}
%
%

The first observable of interest is the kinetic energy of the electrons
defined as
\begin{equation}
  \Ek 
  =
  -t\sum_{\las ij\ras\si}\las\tilde{c}^\dag_{i\si}\tilde{c}^{\nag}_{j\si}\ras
  =
  -2t \sum_{\las ij\ras}
  \las
  c^\dag_{i\UP} c^{\nag}_{j\UP} e^{\rmi\gamma(\op_i-\op_j)}
  \ras
  \,,
\end{equation}
where we have exploited spin symmetry. Following the same steps as in the
derivation of the partition function we get
\begin{equation}
  \las\tilde{c}^\dag_{i\UP}\tilde{c}^{\nag}_{j\UP}\ras
  =
  \Z_L^{-1}\int\,\D p\,w_\text{b}
   e^{\rmi\gamma(p_{i,1}-p_{j,1})}
  \tr_\text{f}
  (
  \Omega\, c^\dag_{i\UP}c^{\nag}_{j\UP}  
  )
  \,.
\end{equation}
Writing out explicitly the fermionic trace we obtain
\begin{eqnarray}\nonumber
  \tr_\text{f}(\Omega\, c^\dag_{i\UP}c^{\nag}_{j\UP})
  &=&
  \sum_{i'j'}
  \bra{i',j'} \Omega \;c^\dag_{i\UP}c^{\nag}_{j\UP}\ket{i',j'}
  \\
  &=&
  \sum_{j'}
  \bra{j,j'}\Omega\ket{i,j'}
  \,,
\end{eqnarray}
and the kinetic energy finally becomes
\begin{equation}
  \Ek 
  =
  -2t\Z_L^{-1}\int\D p w_\text{b}
  \sum_{\las ij\ras}\sum_{j'}
  e^{\rmi\gamma(p_{i,1}-p_{j,1})}
  \bra{j,j'}\Omega\ket{i,j'}
  .
\end{equation}
In addition to $\Ek$, we shall also consider the correlation function
\begin{equation}\label{eq:rho}
  \rho(\delta)
  =
  \sum_i \las \on_{i\UP} \on_{i+\delta\DO}\ras
  \quad
  ,
  \quad
  \delta = 0,1,\dots,N/2-1
  \,.
\end{equation}
A simple calculation leads to
\begin{equation}
  \rho(\delta)
  =
  \Z_L^{-1}\int\,\D p\,w_\text{b} \sum_i
  \bra{i,i+\delta}\Omega\ket{i,i+\delta}
  \,.
\end{equation}
Finally, we would like to point out that other observables, such as the total
energy and the momentum distribution $\las c^\dag_{k\si} c^{\nag}_{k\si} \ras$, may
also be measured within the current approach, while correlation functions
such as $\las\on_i\ox_j\ras$ or the quasiparticle weight cannot be determined
accurately.\cite{HoEvvdL03}

%
%
\subsection{Principal components and reweighting}\label{sec:princ-comp-rewe}
%
%

We make use of the principal component representation and the reweighting
procedure, which have been discussed in detail in I. Defining the
principal components $\bm{\xi}_i=A^{1/2} \bm{p}_i$, in terms of which
$S_\text{b}$ [Eq.~(\ref{eq:action-w-matrix})] takes a Gaussian form which can
be sampled exactly,\cite{HoEvvdL03} allows to perform calculations that
are free of any autocorrelations between successive phonon configurations. In
combination with the reweighting, every new phonon configuration is accepted,
and measurements can be made after each sweep through the $N\times L$
space-time lattice.  The reweighting refers to the use of the purely bosonic
weight $\wb$ in the QMC simulation, while all the influence of the electrons
and their interaction with the phonons---contained in $\wf$---is treated
exactly as part of the observables.

%
%
\subsection{Numerical details and performance}\label{sec:numerical-details}
%
%

The most significant difference between the present calculations and the
one-electron case in I is the dimension of the matrices involved. While for one electron all matrices
have size $N\times N$---$N$ being the extension  of the 1D lattice under
consideration---here the dimension is $N^2\times N^2$.
Clearly, this restricts calculations with respect to the number of lattice
sites, especially in higher dimensions $\rD>1$ where $N^2\mapsto N^{2\rD}$. The total
numerical effort for the current approach is proportional to $N^{6\rD} L$. In
contrast, the one-electron algorithm\cite{HoEvvdL03} displays the same dependence $\propto
N^{3\rD}L$ as the determinant QMC method of Blankenbecler
\etal\cite{BlScSu81} for the many-electron case, which can be reduced to
$N^{2\rD}L$ by employing the checkerboard breakup of the hopping
matrix.\cite{wvl1992} The increase in required
computer time for the bipolaron results from the fixed number of electrons.
Recently, a grand-canonical version of the one-electron algorithm, also with
a computer time $\sim N^{2\rD}L$, has been applied to study the
dependence of polaron formation on carrier density in the spinless Holstein
model.\cite{HoNevdLWeLoFe04}
For the bipolaron problem, we shall see below
that the present algorithm allows one to study lattices of reasonable size
$N\leq14$, for a wide range of the parameters $\omb$, $\lambda$ and $\Ub$. In
particular, we can obtain accurate results in the adiabatic regime $\omb<1$.

Let us briefly compare our method to other QMC approaches to the
HH bipolaron. The method of de Raedt and
Lagendijk\cite{deRaLa86} is based on an analytic integration over the phonon
degrees of freedom, leading to a model with retarded el-el interaction.
Similar to our approach, it employs a Suzuki-Trotter approximation and gives
results at finite temperatures. For simplicity, de Raedt and Lagendijk only
considered the adiabatic limit $\omb=0$, in which there are no retardation
effects. The numerical effort grows as $L^2$, but is virtually independent of
the system size, so that simulations can be carried out even for large
clusters in three dimensions.  However, it is not clear how a small but
finite phonon frequency $\omb<1$ will affect the computer time.

Macridin \etal \cite{Mac04} used the diagrammatic QMC method to study two electrons
on a $25\times25$ lattice. Although their approach does not rely on the Suzuki-Trotter decomposition, it
is limited to zero temperature, and statistical errors increase noticeably for
$\omb<1$.  Moreover, the accuracy also decreases for
large values of $\lambda$ and/or $\Ub$, whereas we shall see in
Sec.~\ref{sec:results} that we can easily study the strong el-ph
coupling regime also for $\Ub>0$.

In I, we announced the possibility of reducing the numerical effort for the
present method by exploiting the translational invariance of the model. To
this end, the basis
states (\ref{eq:basis}) would have to be replaced by states $\{\ket{k,\Delta}\}$
with total quasimomentum $k$, and with the two electrons separated by a
distance $\Delta$. A similar idea has been used by
Kornilovitch\cite{Ko98,Ko99} for a single electron. In one dimension,
the use of the basis $\{\ket{k,\Delta}\}$ would reduce the size of the matrices
in the algorithm from $N^2\times N^2$ to $N\times N$. However, in the course
of the simulation, we had to evaluate the matrix product over $\tau$
[Eq.~(\ref{eq:matproduct})] for each
allowed value of $k$. In total, we could therefore reduce the numerical effort by a
factor $N$. 
The major drawback of using the reduced basis in momentum space is that it
significantly complicates the program code. Consequently, in this work,
we have restricted ourselves to the straight-forward extension of the
one-electron algorithm presented in I.

Finally, the minus-sign problem, which has been mentioned in I, also exists
here. However, as for one electron, it quickly diminishes with
increasing system size, and does therefore not conceivably affect
simulations.

%
%
%
\section{\label{sec:VPA}Variational approach}
%
%
%

Although the
method can easily be applied also in higher dimensions, we wish to keep the
notation simple and therefore restrict the derivation to $\rD=1$. The
approximation consists of the use of a zero-phonon basis after the extended
unitary transformation, which leads to $\tilde{I}_\text{ep}=0$
[Eq.~(\ref{eq:htilde})]. Furthermore, neglecting the ground-state
energy of the oscillators, we also have $P=0$, so that
\begin{equation}\label{eq:htilde_vpa}
  \tilde{H}
  =
  \tilde{K}+\tilde{I}_\text{ee}\,,
\end{equation}
with the transformed hopping term
\begin{equation}
  \tilde{K}
  =
  -t_\text{eff}\sum_{\las ij\ras\si}
  c^\dag_{i\si}c^{\nag}_{j\si}
  =
  \sum_{k\si} \varepsilon(k)\; c^\dag_{k\si}c^{\nag }_{k\si}
\end{equation}
and $\varepsilon(k)=- 2\; t_\text{eff} \sum_{k\si} \cos(k)$.
Here, the effective hopping amplitude is given by\cite{HoEvvdL03}
\begin{equation}\label{eq:t_eff}
  t_\text{eff}
  =
  \frac{1}{z}\sum_\delta
  e^{-\frac{1}{4}\sum_l(\gamma_{l-\delta}-\gamma_{l})^2}
  \,,
\end{equation}
where $\delta=\pm 1$ in one dimension, $z$ is the
number of nearest neighbors, and rotational invariance has been
exploited. For two electrons of opposite spin, the interaction
term~(\ref{eq:htilde_el-el}) simplifies to
\begin{equation}
  \tilde{I}_\text{ee}
  =
  2 v_0 - U + 2 \sum_{ij} v_{ij} \on_{i\UP} \on_{j\DO}
\end{equation}
if we use $v_{ij}=v_{|j-i|}$ and $\on_{i\sigma}\on_{j\sigma}=0$ for
$i\ne j$. The two-electron eigenstates of the
Hamiltonian~(\ref{eq:htilde_vpa}) have the form
\begin{equation}\label{eq:states_k}
  \ket{\psi_k}
  =
  \sum_p  \tilde d^{\nag}_p c^\dag_{k-p\DO} c^\dag_{p\UP}\ket{0}
  \,.
\end{equation}
Here we have suppressed the phonon component which is simply given by the
ground state of $N$ free harmonic oscillators. The
states~(\ref{eq:states_k}) may be written as
\begin{equation}\label{eq_psi_rs}
  \ket{\psi_k}
  =
  \frac{1}{\sqrt{N}}\sum_i e^{\rmi k x_i}
  \sum_l d^{\nag}_l c^\dag_{i\DO} c^\dag_{i+l\UP}\ket{0}
  \,,
\end{equation}
where the Fourier transform
\begin{equation}\label{eq:FT}
  \bm{d}
  =
  F \tilde{\bm{d}}
\end{equation}
with $F_{lp} = e^{\rmi  x_l p}/\sqrt{N}$ has been employed.
The normalization of Eq.~(\ref{eq:states_k}) reads
$\bra{\psi_k}\psi_k\ras =\sum_p |d_p|^2$.

The expectation value of the transformed hopping term with respect to the
states defined by Eq.~(\ref{eq:states_k}) becomes
\begin{widetext}
\begin{eqnarray}\nonumber
  \bra{\psi_k}\tilde{K}\ket{\psi_k}
   &=&
   \sum_{pp'}
   \tilde d_p^* \tilde d_{p'}^{\phantom{*}}
   \sum_{q} \varepsilon(q)
   \bigg(
   \underbrace{%
     \bra{0}
     c^{\nag }_{p\UP} c^{\nag}_{k-p\DO}
     \on^{\nag }_{q\UP}
     c^{\dag }_{k-p'\DO} c^\dag_{p'\UP}\ket{0}
     }
     _{\delta_{p,p'}\delta_{q,p}}
   +
   \underbrace{%
     \bra{0}
     c^{\nag }_{p\UP} c^{\nag}_{k-p\DO}
     \on^{\nag }_{q\DO}
     c^{\dag }_{k-p'\DO} c^\dag_{p'\UP}\ket{0}
     }
     _{\delta_{p,p'}\delta_{q,k-p}}
   \bigg)
   \\\nonumber
   &=&
   \sum_p |\tilde d_p|^2 \left[\varepsilon(p) + \varepsilon(k-p)\right]
   =
   -4\;t_\text{eff}\,\bm{d}^\dag T_k \bm{d}
   \,.
\end{eqnarray}
In the last step we introduced vector notation, defined $T_k=F
\,\text{diag}[\cos (p) + \cos (k-p)\big]/2\,F^\dag$ and used Eq.~(\ref{eq:FT}).
The expectation value of the interaction term is best computed in the
real-space representation~(\ref{eq_psi_rs}). We find
\begin{eqnarray}\nonumber
  \bra{\psi_k}{\tilde I}_\text{ee}\ket{\psi_k}
  &=&
  (2v_0-U)\sum_l |d_l|^2
  + \frac{2}{N} \sum_{ij} v_{ij}
  \sum_{j'j''}\sum_{ll'}
  d_l^* d^{\phantom{*}}_{l'}
   e^{\rmi k (x_{l\phantom{'}}-x_{l'})}
  \underbrace{
    \bra{0}
    c^{\nag}_{j' + l\UP}  c^{\nag}_{j'\DO}
    \on^{\nag}_{i\UP} \on^{\nag}_{j\DO}
    c^\dag_{j''\DO} c^\dag_{j'' + l'\UP}
    \ket{0}}_{
    \delta_{jj'}\delta_{jj''}\delta_{i,j+l}\delta_{l,l'}
    }
  \\\nonumber
  &=&
  (2v_0-U) \sum_l |d_l|^2
  + \frac{2}{N} \sum_{j,l} v_{j+l,j} |d_{l}|^2
  =
  (2v_0-U) \bm{d}^\dag \bm{d} + 2\bm{d}^\dag V \bm{d}
  \,,
\end{eqnarray}
\end{widetext}
where the diagonal matrix $V_{ij}=\delta_{ij} v_{i}$ has been introduced.
The minimization with respect to $\bm{d}$ yields the eigenvalue problem
\begin{equation}\label{eq:minimize}
  (-4t_\text{eff} \;T_k + 2 V)\,\bm{d}
  =
  (E_0 - 2v_0 + U)\,\bm{d}
  \,.
\end{equation}
The vector of coefficients $\bm{d}$ and thereby the ground state are
determined by minimizing the ground-state energy $E_0$ through variation of
the displacement fields $\gamma_{ij}$. 
In the present work, we use the
unconstrained nonlinear optimization routine {\it fminsearch} from the MATLAB
package, together with several different starting points, including the
simple LF result and random values of the $\gamma_{ij}$. This ensures
reproducible results even for a large number of variational parameters.

In contrast to the local LF
transformation, this procedure takes into account displacements of the
oscillators not only at the same but also at the sites surrounding the two
electrons. This represents a physically much better ansatz to describe the
extended state which exists for weak el-ph coupling and/or strong Coulomb
repulsion. Similar to the one-electron problem, we shall refer to the result
obtained from the above variational method by replacing $\gamma_{ij}$ with
$\gamma\delta_{ij}$ as the Holstein Lang-Firsov (HLF) approximation.

%
%
%
\section{\label{sec:results}Results}
%
%
%

Before we turn to the results, we would like to review briefly the physics of
the one-dimensional HH bipolaron as it emerges from existing work (see
Sec.~\ref{sec:holstein}). In the absence of Coulomb repulsion, the two
electrons form a bound state for any $\lambda>0$. A crossover from an
extended state, also called a {\it large bipolaron}, to a {\it small
  bipolaron}---with both electrons occupying the same site---is observed at a
critical coupling strength $\lambda_\text{c}$.  The value of
$\lambda_\text{c}$ is determined by the competition between the different
terms in the Hamiltonian~(\ref{eq:holstein}). Similar to the one-electron
case, for small phonon frequencies, the crossover takes place when the gain
in potential energy due to bipolaron formation overcomes the loss in kinetic
energy. While the former can be estimated in the atomic limit as $4\Ep$ (see,
\eg, Ref.~\onlinecite{HoAivdL04}), the latter is given here by $W=4t$---the
kinetic energy of the two electrons at $\lambda=0$. Since $\lambda$ can also be
written in the form $\lambda=\frac{1}{2}(4\Ep/W)$, we expect
$\lambda_\text{c}=0.5$. For larger phonon frequencies $\omb\gg1$, the lattice
energy plays an important role, and gives rise to the additional criterion
$2\sqrt{\Ep/\om}>1$ for the existence of a small bipolaron.\cite{ChRaFe99}

For $\Ub>0$, a state with two weakly bound polarons is stable for weak enough
el-ph interaction. Interestingly, starting from a small bipolaron a cross
over to an {\it intersite bipolaron}---with the two electrons being
localized most likely at neighboring lattice sites---takes place at a
critical value $\Ub_\text{c}$.\cite{LaMaPu97,BoKaTr00,WeFeWeBi00}  This
state has been shown to have a much smaller effective mass than an on-site
bipolaron,\cite{BoKaTr00} and may therefore exist as a mobile quasiparticle
in real systems. Phase diagrams of the intersite bipolaron
have been reported in one\cite{BoKaTr00,WeFeWeBi00} and two
dimensions.\cite{Mac04} For $\omb=1$, the region where such a state exists in
one dimension is quite accurately described by $U<2\Ep$ at weak el-ph
coupling, and by $U<4\Ep$ at strong coupling.\cite{BoKaTr00}

If $2\Ep\gg U$ the two electrons can overcome the on-site repulsion and
form a small bipolaron. As we shall see below, the competition between
Hubbard repulsion and attractive interaction due to the electron-lattice
coupling depends critically on the phonon frequency.  A summary of the
different bipolaron states together with (approximate) conditions for their
existence is given in Tab.~\ref{tab:bipolaron:bipolaronconditions}.
\begin{table}
  \centering
  \caption{\label{tab:bipolaron:bipolaronconditions}Conditions for the
    existence of different singlet bipolaron states (see text). Here ``wc'' and
    ``sc'' denote weak coupling and strong coupling, respectively.}\vspace*{1em}
  \begin{tabular}{cc|cc}\hline\hline
     \multicolumn{4}{c}{$U=0$}                 \\\hline
      Large bipolaron     &\quad\,&\quad\,&   Small bipolaron \\\hline
      $\lambda<0.5$       &&&   $\lambda>0.5$\\
      or                  &&&   and          \\
      $2\sqrt{\Ep/\om}<1$ &&&   $2\sqrt{\Ep/\om}>1$\\\hline
    \end{tabular}
    \vspace*{1em}
    \begin{tabular}{c|c|c}
       \multicolumn{3}{c}{$U>0$}                      \\\hline
       Two                 & Intersite      &   Small \\
       polarons            & bipolaron      &   bipolaron\\\hline
       $U>2\Ep$ (wc)       &  $U<2\Ep$ (wc) &   \\
       $U>4\Ep$ (sc)       &  $U<4\Ep$ (sc) &   \raisebox{1.5ex}[-1.5ex]{$U\ll2\Ep$}\\
       \hline\hline
  \end{tabular}
\end{table}

We would like to mention that very interesting physics can also be deduced
from the bipolaron band dispersion (see Ref.~\onlinecite{HoAivdL04} and
references therein).  Although the latter can be studied in principle within
our variational approach, the subtle effects originating from the retarded
nature of the el-ph interaction\cite{HoAivdL04} could not be
addressed in a satisfactory way.

%
%
\subsection{Quantum Monte Carlo}\label{sec:res_qmc}
%
%

To eliminate the error $\sim(\dtau)^2$ due to the Suzuki-Trotter approximation,
we extrapolate the QMC results to $\dtau=0$. In contrast, this error is
expected to be relatively large (on the order of a few percent) in the
calculations of de Raedt and Lagendijk, due to the use of a rather small
number of Trotter slices ($L=32$ at $\beta=5$, so that $\dtau\approx0.16$;
see Ref.~\onlinecite{deRaLa86}). Here we have performed simulations for three
different values of $\dtau$, typically 0.1, 0.075 and 0.05. The errorbars in
the figures below are usually as small as the linewidth, and will not be
shown if smaller than the symbols used.

\begin{figure}
  \includegraphics[width=0.45\textwidth]{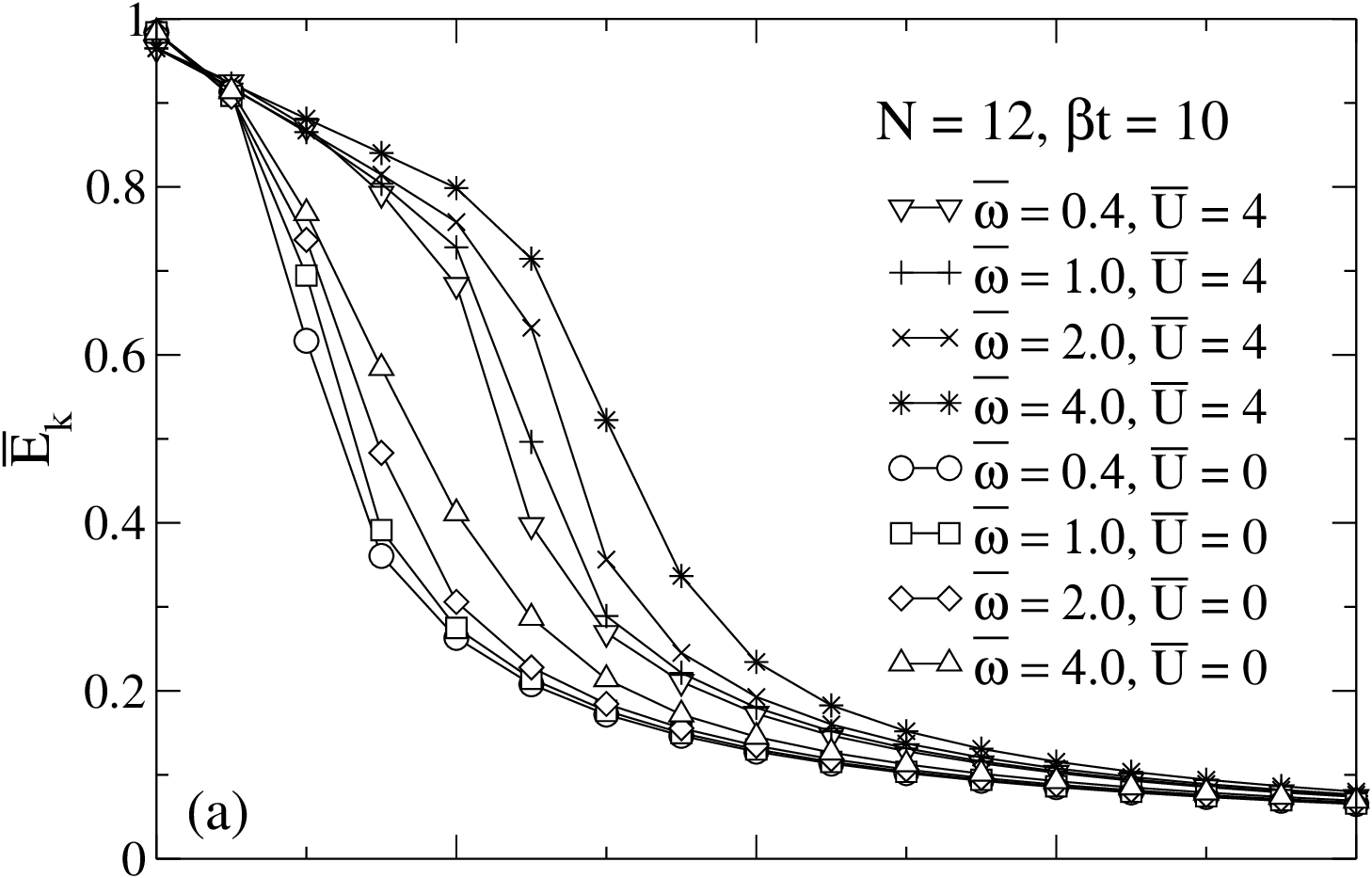}\\
  \includegraphics[width=0.45\textwidth]{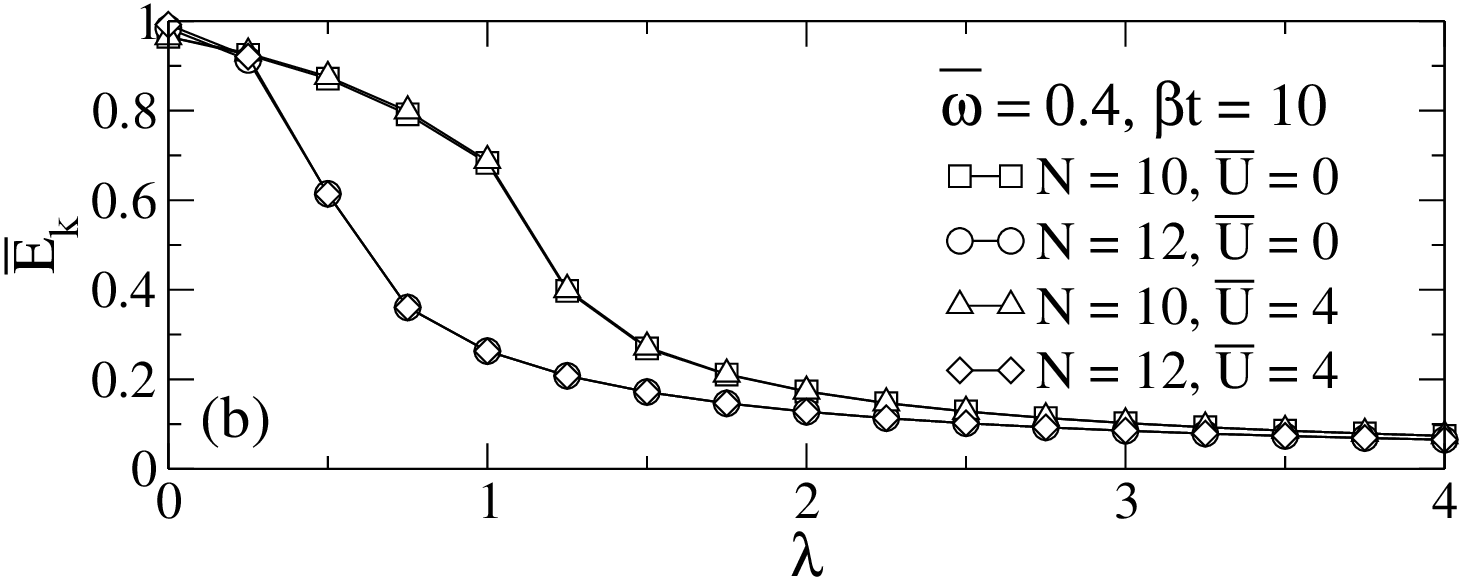}
 \caption{\label{fig:Ek_lambda_omega}
   (a) Normalized kinetic energy $\Ekb$ [Eq.~(\ref{eq:ekeff})] from QMC as a
   function of the el-ph coupling $\lambda$ for different values of
   the phonon frequency $\omb$ and the Hubbard repulsion $\Ub$.
   (b) Dependence of $\Ekb$ on the cluster size $N$.
   Here and in subsequent figures QMC results have been extrapolated to $\dtau=0$ (see
   text), errorbars are suppressed if smaller than the symbol size, and lines are guides to
   the eye.  }
\end{figure}

Owing to the increased numerical effort compared to the one-electron
case,\cite{HoEvvdL03} we shall only present results for $N\leq12$.
Fortunately,
finite-size effects on, \eg, the kinetic energy, are already very small
for this cluster size, as illustrated by Fig.~\ref{fig:Ek_lambda_omega}(b)
for the most critical parameters. As expected, the largest changes with $N$
occur near the crossover to a small bipolaron.\cite{HoEvvdL03} Similar
behavior has been found for the correlation function $\rho(\delta)$.

We define the effective kinetic energy of the two electrons as
\begin{equation}\label{eq:ekeff}
  \Ekb
  =
  \Ek/(-4t)
  \,.
\end{equation}
In Fig.~\ref{fig:Ek_lambda_omega}(a) we depict $\Ekb$ as a function of the el-ph
coupling for different values of $\omb$ and $\Ub$, at $\beta t=10$. 
This value of the inverse temperature is twice as large as in previous
work,\cite{deRaLa86} yielding results sufficiently close to the ground state 
to reveal the effects of bipolaron formation.

Figure~\ref{fig:Ek_lambda_omega}(a) reveals a strong decrease of $\Ekb$ near
$\lambda=0.5$ for small phonon frequencies and $\Ub=0$. With increasing
$\omb$, the crossover becomes less pronounced, and shifts to larger values
of $\lambda$.  For the same value of $\omb$, the crossover to a small
bipolaron is sharper than the small-polaron crossover in
the Holstein model with one electron (see, \eg, I).
The small but finite kinetic energy even for strong el-ph interaction is a
result of undirected, internal motion of the two electrons inside the phonon
cloud.  For a finite on-site repulsion $\Ub=4$, $\Ekb$
remains fairly large up to $\lambda\approx1$ (for $\omb\lesssim2.0$), in
agreement with the strong-coupling result $\lambda_\text{c}=1$ for $\Ub=4$
(see discussion in Ref.~\onlinecite{HoAivdL04}).  At even stronger coupling,
the Hubbard repulsion is overcome, and a small bipolaron is formed. Again, we
see that the critical coupling increases with increasing phonon frequency.
We would like to mention that the kinetic energy has also been calculated by
ED on clusters of up to twelve sites,\cite{WeRoFe96,FeRoWeMi95,WeFeWeBi00}
but results for $\omb<1$ were restricted in the accessible range
of $\lambda$. In the regime where ED is applicable, a very good agreement has
been found with our QMC data.

The nature of the bipolaron state is revealed by the correlation function
$\rho(\delta)$ defined in Eq.~(\ref{eq:rho}), which gives the
probability for the two electrons to be separated by a distance $\delta\geq0$, and
therefore represents a direct measure for the size of the bipolaron. Clearly,
we have the sum rule $\sum_\delta \rho(\delta)=1$. As pointed out, \eg, by
Marsiglio,\cite{Marsiglio95} the phonon frequency determines the degree of
retardation of the el-ph interaction, and thereby sets the maximal allowed
distance between the two electrons compatible with a bound state. In the
sequel, we shall focus on the most interesting case of small phonon
frequencies, which has often been avoided in previous work for reasons
outlined in Sec.~\ref{sec:holstein}.

\begin{figure}
  \includegraphics[width=0.45\textwidth]{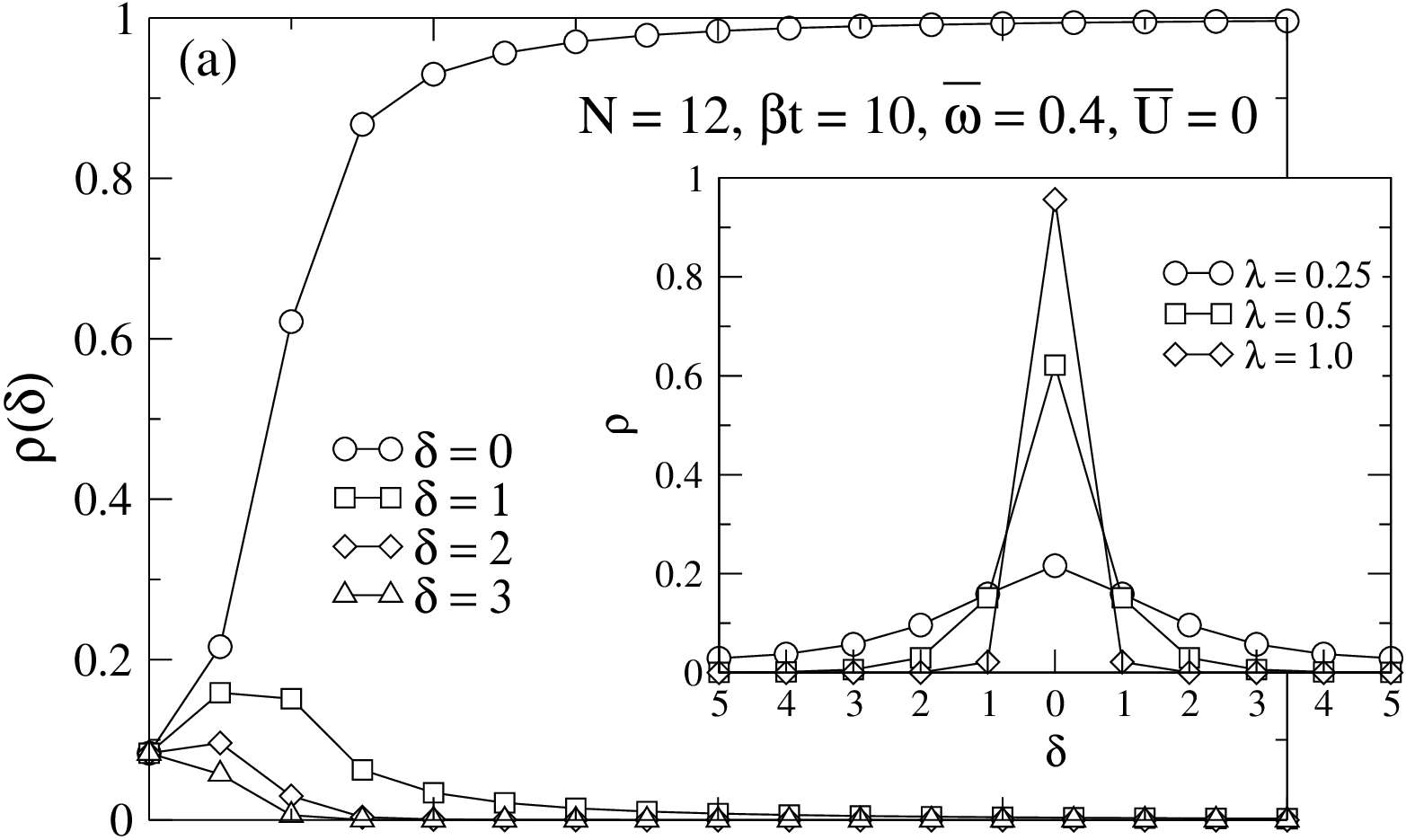}\\
  \includegraphics[width=0.45\textwidth]{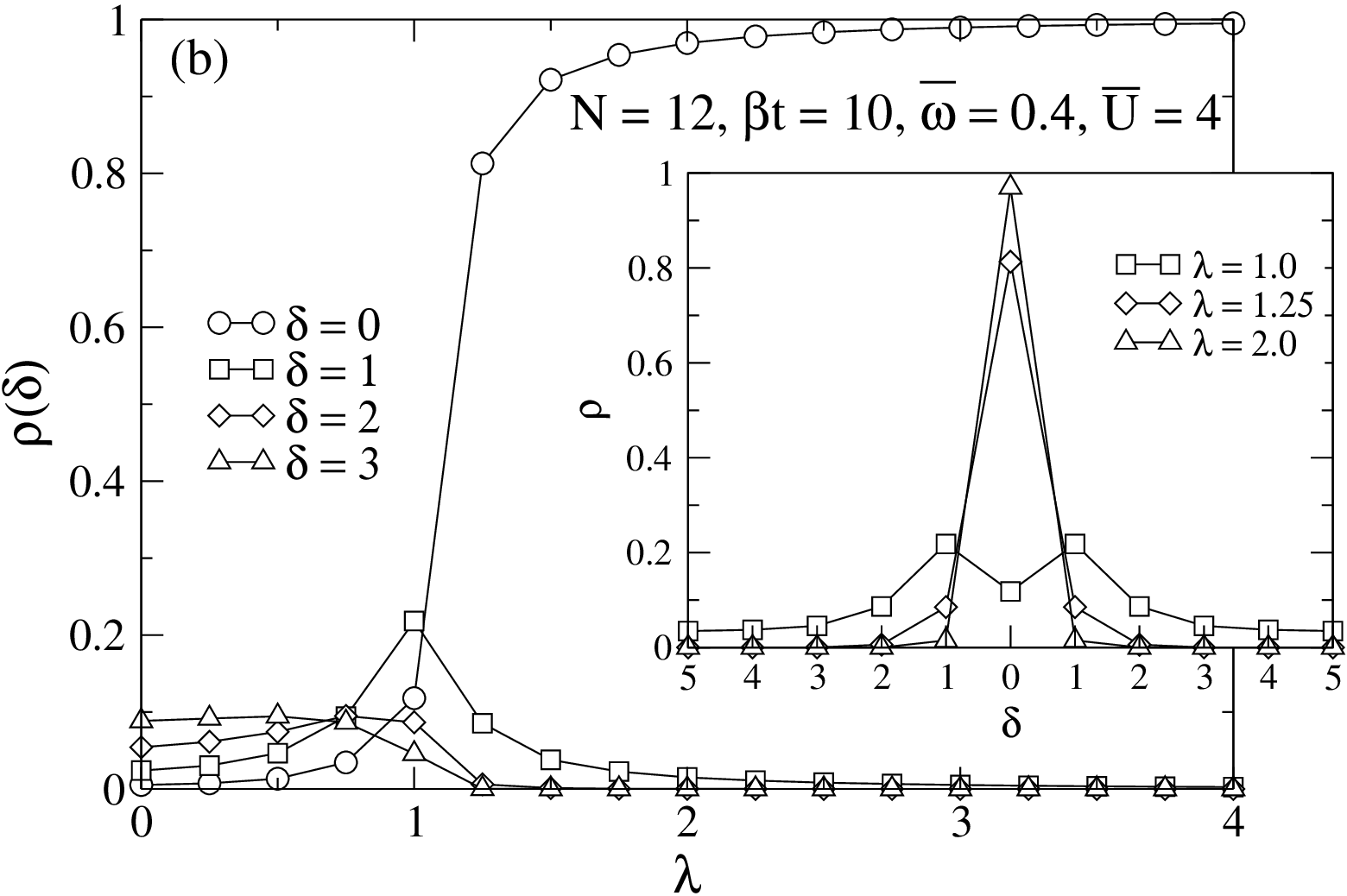}
 \caption{\label{fig:rho}
   Correlation function $\rho(\delta)$ [Eq.~(\ref{eq:rho})] from QMC as a function of
   el-ph coupling $\lambda$ for different values of $\delta$.  [(a)
   $\Ub=0$, (b) $\Ub=4$]. Inset: Correlation function $\rho(\delta)$ as a
   function of the distance $\delta$ of the electrons for different $\lambda$.}
\end{figure}

Figure~\ref{fig:rho}(a) shows $\rho(\delta)$ as a function of $\lambda$ for
$\Ub=0$. Starting from the noninteracting state ($\lambda=0$) with
$\rho=1/N$, we see a pronounced increase of $\rho(0)$ near $\lambda=0.5$. For
large $\lambda\gtrsim2$, we have $\rho(0)\approx1$ and $\rho(\delta)\approx0$
for $\delta>0$, characteristic for the on-site
bipolaron. The decrease of the spatial extent of the bipolaron with
increasing el-ph interaction is better illustrated in the inset of
Fig.~\ref{fig:rho}(a), where we depict $\rho$ as a function of $\delta$. For
finite on-site repulsion $\Ub=4$, an extended bipolaron state is stabilized
for small $\lambda$ [Fig.~\ref{fig:rho}(b)], while a small bipolaron is found
for $\lambda=2$.  Additionally, we see that for $\lambda=1$, the electrons
are most likely to occupy neighboring lattice sites [intersite bipolaron, see
also inset in Fig.~\ref{fig:rho}(b)].

\begin{figure}
  \includegraphics[width=0.45\textwidth]{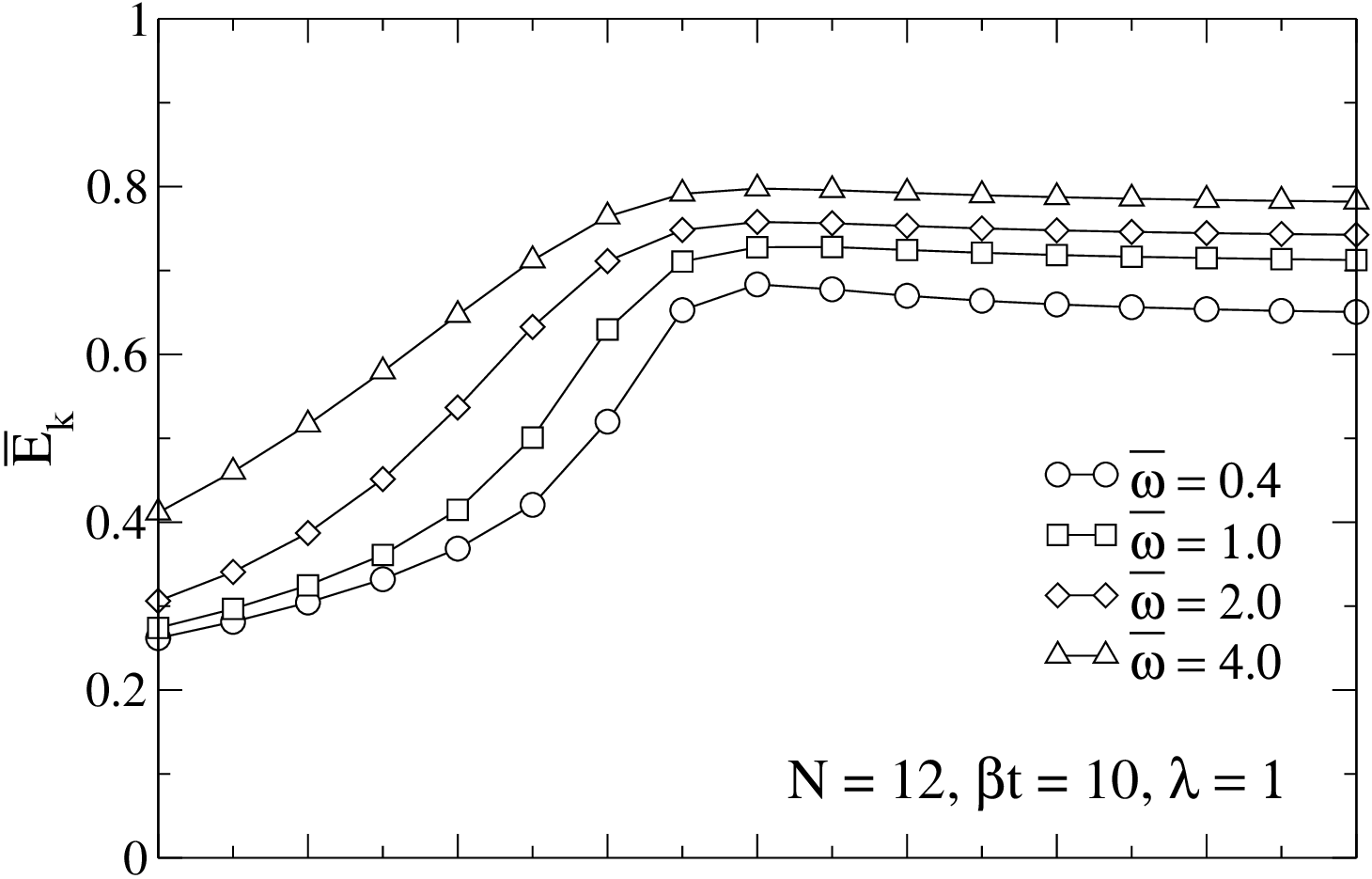}\\
  \includegraphics[width=0.45\textwidth]{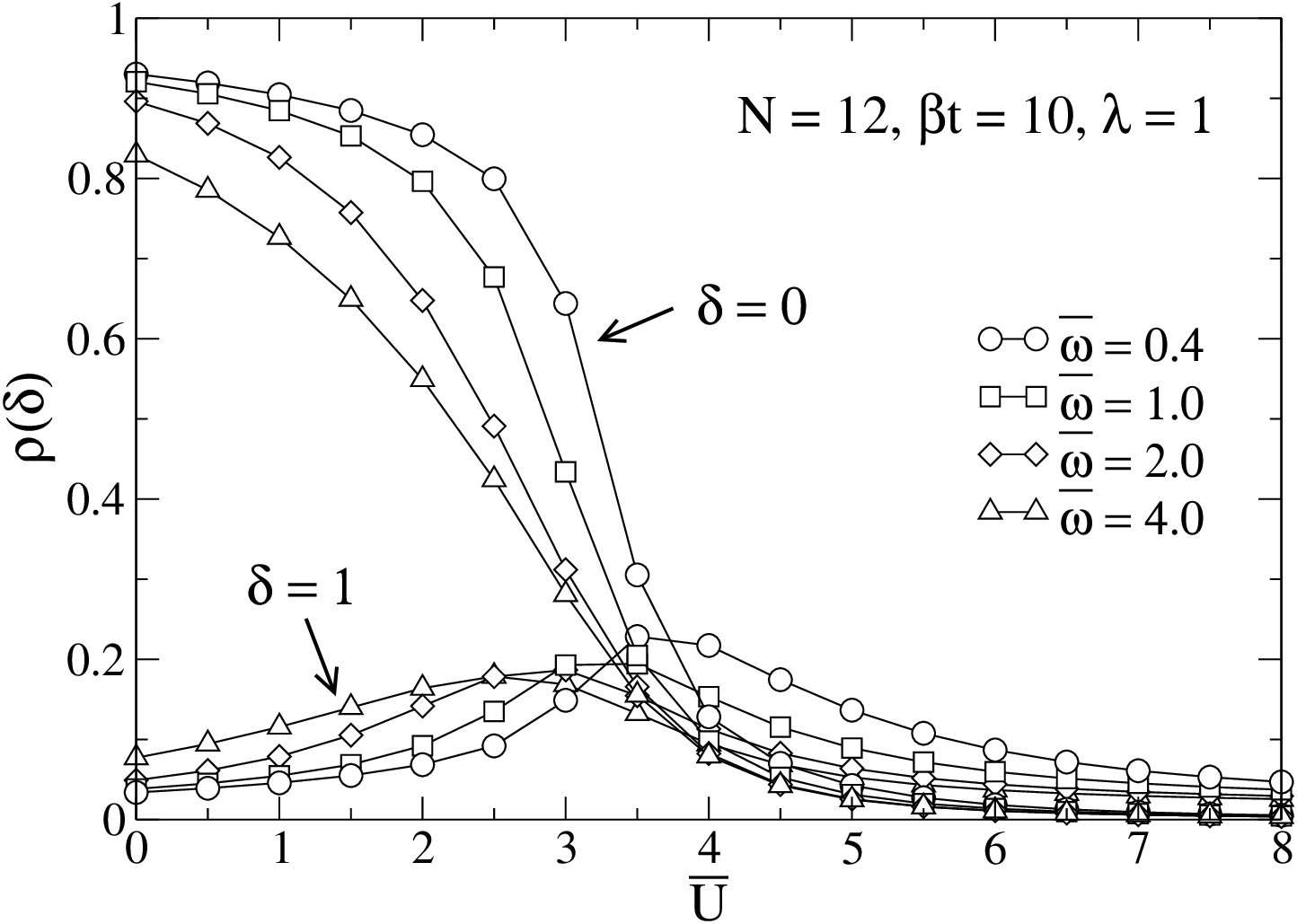}
 \caption{\label{fig:S0S1}
   (a) Normalized kinetic energy $\Ekb$ and (b) correlation functions
   $\rho(0)$, $\rho(1)$ from QMC as a function of the Hubbard repulsion $\Ub$ for
   different values of the phonon frequency $\omb$.}
\end{figure}

As pointed out earlier, a crossover from a small to an intersite bipolaron to two
weakly bound polarons takes place as a function of the Hubbard interaction.
Since the latter competes with the retarded el-ph interaction,
the phonon frequency is expected to be an important parameter. In
Fig.~\ref{fig:S0S1}, we show the kinetic energy and the correlation function
$\rho(\delta)$ as a function of $\Ub$. We have fixed the el-ph coupling to
$\lambda=1$. Starting from a small bipolaron for $\Ub=0$ [see
Fig.~\ref{fig:rho}(a)], the kinetic energy increases with increasing Hubbard
repulsion, equivalent to a reduction of the effective bipolaron
mass.\cite{BoKaTr00,ElShBoKuTr03}
Although the crossover is slightly washed out by the finite
temperature in our simulations, there is a well-conceivable increase in
$\Ekb$ up to $\Ub\approx4$, above which the kinetic energy begins to decrease
again. The increase of $\Ekb$ originates from the breakup of the small
bipolaron, as indicated by the decrease of $\rho(0)$ in
Fig.~\ref{fig:S0S1}(b). Close to $\Ub=4$, the curves for $\rho(0)$
and $\rho(1)$ cross, and it becomes more favorable for the two electrons to
reside on neighboring sites.  The intersite
bipolaron only exists below a critical Hubbard repulsion $U_\text{c}$. As
discussed at the beginning of this section, the latter is given by
$U_\text{c}=2\Ep$ (\ie, here $\Ub_\text{c}=4$) at weak el-ph
coupling, and by $U_\text{c}=4\Ep$ at strong coupling. For an intermediate
value $\lambda=1$ as in Fig.~\ref{fig:S0S1}, the crossover from the
intersite state to two weakly bound polarons is expected to occur somewhere
in between, but is difficult to identify exactly from the QMC results.

Figure~\ref{fig:S0S1} further illustrates that the crossover becomes steeper with
decreasing phonon frequency. In the adiabatic limit $\omb=0$, it has been
shown to be a first-order phase transition,\cite{PrAu99} whereas for $\omb>0$
retardation effects suppress any nonanalytic behavior. At the same
$\Ub$, $\Ekb$ increases with $\omb$ since for a fixed $\lambda$, the
bipolaron becomes more weakly bound. For the same reason, the crossover to an intersite
bipolaron---showing up in Fig.~\ref{fig:S0S1} as a crossing of $\rho(0)$ and
$\rho(1)$---shifts to smaller values of $\Ub$.

\begin{figure}
  \includegraphics[width=0.45\textwidth]{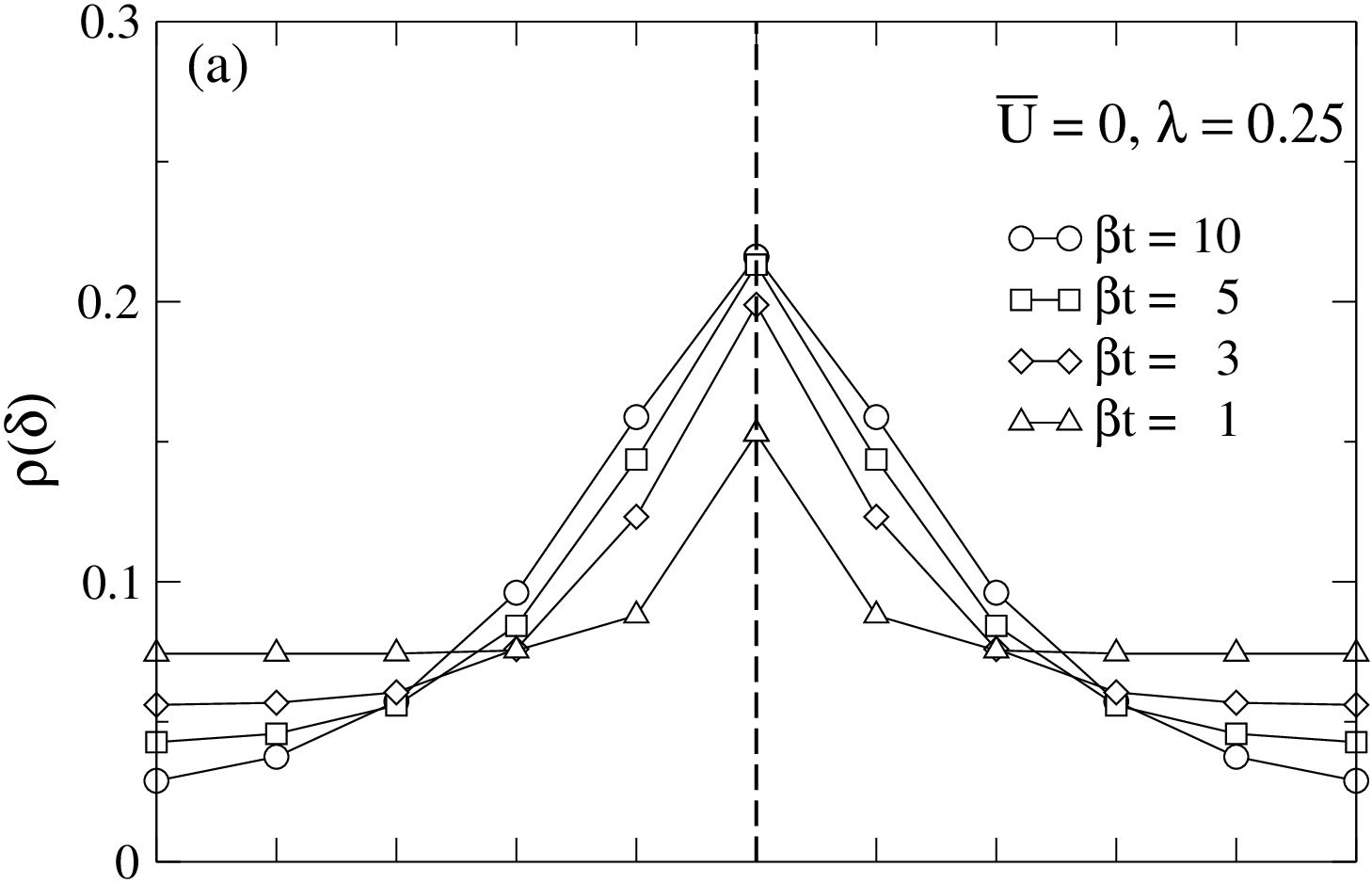}\\
  \includegraphics[width=0.45\textwidth]{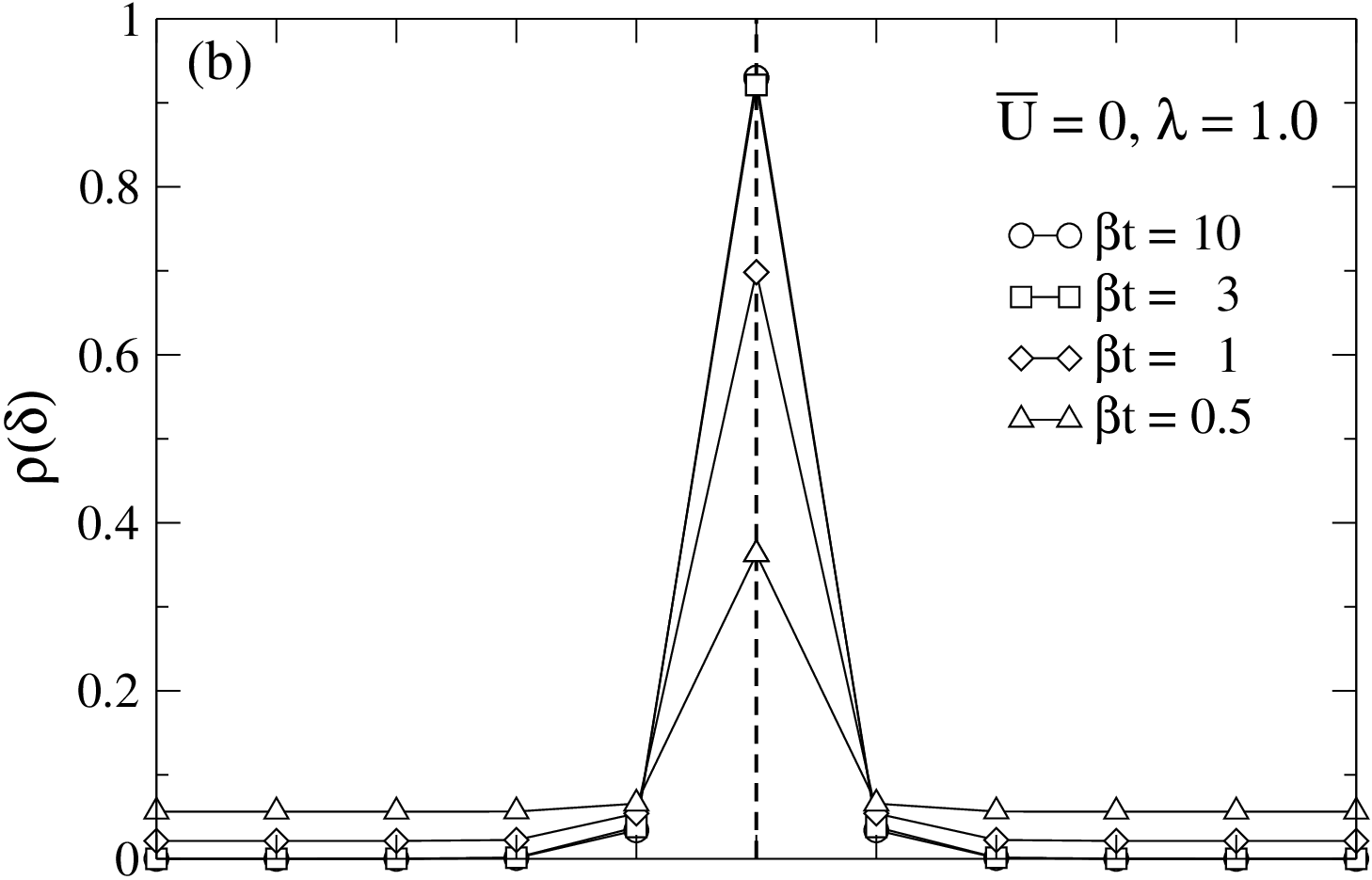}\\
  \includegraphics[width=0.45\textwidth]{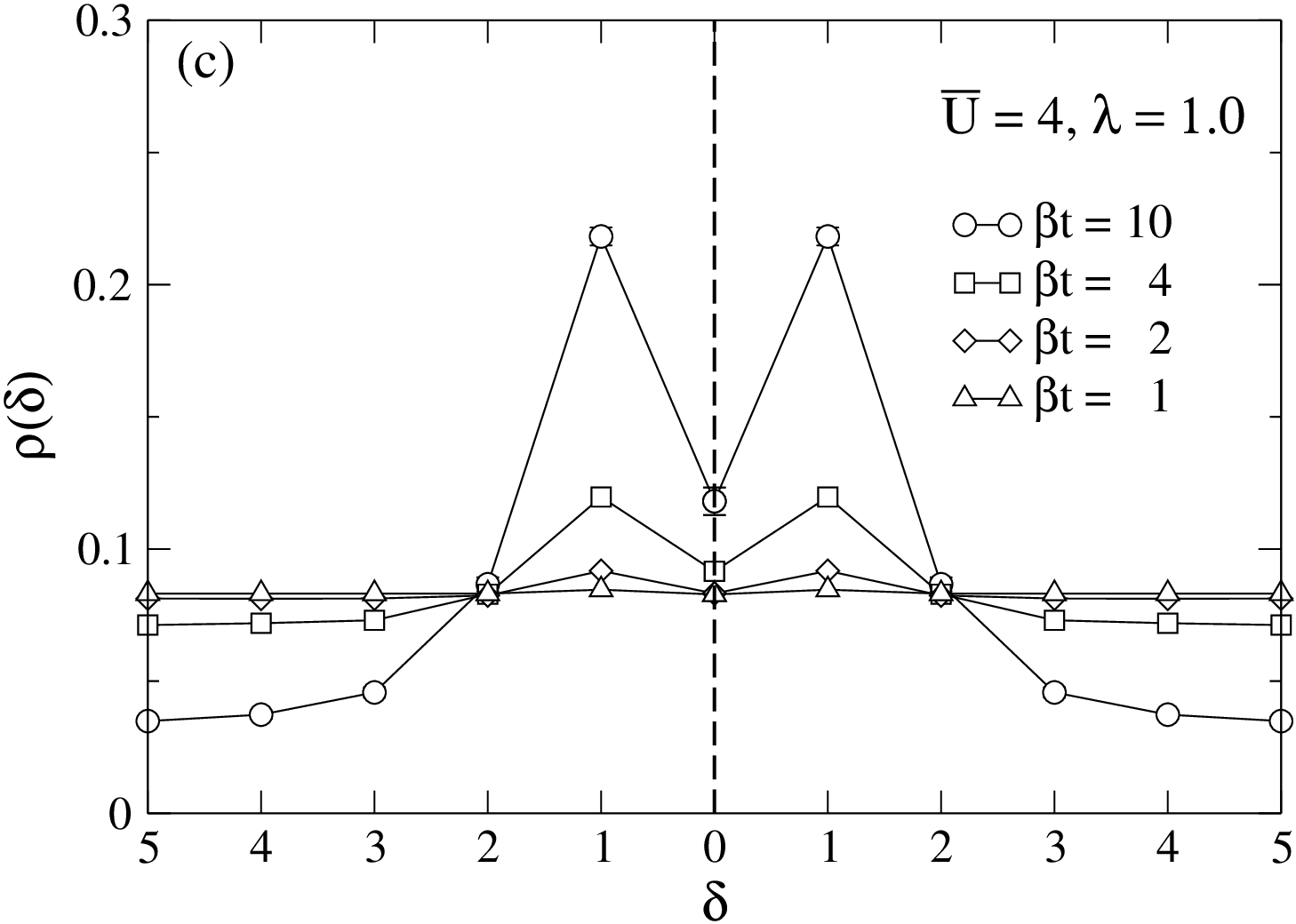}
  \caption{\label{fig:rho_temp}
    Correlation function $\rho(\delta)$ from QMC as a function of $\delta$ for
    different inverse temperatures $\beta$, $N=12$ and $\omb=0.4$.}
\end{figure}

Let us now consider the effect of temperature. While the kinetic energy shows
a similar dependence as in the one-electron case---with the crossover being
smeared out at high temperatures---it is much more interesting to look at
$\rho(\delta)$. In Figs.~\ref{fig:rho_temp}(a)\,--\,(c) we plot
$\rho(\delta)$ at different temperatures, for parameters corresponding to the
three regimes of a large, small and intersite bipolaron, respectively.

\paragraph*{Large bipolaron.}
For the parameters chosen ($\Ub=0$, $\lambda=0.25$), the two electrons are
most likely to occupy the same site, but
the bipolaron extends over a distance of several lattice constants
[Fig.~\ref{fig:rho_temp}(a)]. Clearly, in this
regime, the cluster size $N=12$ used here is not completely satisfactory, but
still provides a fairly accurate description as can be deduced from
calculations for $N=14$ (not shown). Nevertheless, on such a small cluster,
no clear distinction between an extended bipolaron and two weakly bound
polarons can be made.  As the temperature increases from $\beta t=10$ to
$\beta t = 1$, the probability distribution broadens noticeably, so that it
becomes more likely for the two electrons to be further apart. In particular,
for the highest temperature shown, $\rho(0)$ has reduced by about 30 \%
compared to $\beta t=10$.

\paragraph*{Small bipolaron.}
A different behavior is found for the small bipolaron, which exist at
stronger el-ph coupling $\lambda=1.0$. Figure~\ref{fig:rho_temp}(b) reveals
that $\rho(\delta)$ peaks strongly at $\delta=0$, while it is very small for
$\delta>0$ at low temperatures. Increasing the temperature, we observe that
$\rho(\delta)$ remains virtually unchanged up to $\beta t = 3$. Only at very
high temperatures there occurs a noticeable transfer of probability from
$\delta=0$ to $\delta>0$. At the highest temperature shown, $\beta t=0.5$, the two
electrons have a nonnegligible probability for traveling a finite distance
$\delta>0$ apart, although most of the probability is still contained in the
peak located at $\delta=0$.

\paragraph*{Intersite bipolaron.}
Finally, we consider in Fig.~\ref{fig:rho_temp}(c) the intersite bipolaron,
which has been found above for $\Ub=4$ and $\lambda=1.0$
[Fig.~\ref{fig:rho}(b)]. At low temperatures, $\rho(\delta)$ takes on a
maximum for $\delta=1$. For smaller values of $\beta t$, the latter
diminishes, until at $\beta t=1$, the distribution is completely flat, so
that all $\delta$ are equally likely.

The different sensitivity of the bipolaron states to changes in temperature
found above can be explained by their different binding energies. The latter
is given by $\Delta E_0=E_0^{(2)}-2 E_0^{(1)}$, where $E_0^{(1)}$
and $E_0^{(2)}$ denote the ground-state energy of the model with one and two
electrons, respectively. These quantities can be calculated using the present
method as well that presented in I.

Generally, the thermal dissociation is expected to occur at a temperature
such that the thermal energy $k_\text{B} T = (\beta T)^{-1}$ becomes
comparable to $\Delta E_0$, in accordance with our numerical data.
The large and the intersite bipolaron are relatively weakly
bound as a result of the rather small effective interaction
$U_\text{eff}\approx U-2\Ep$ (see discussion in Ref.~\onlinecite{HoAivdL04}).
The binding energies are $\Delta E_0\approx-(0.32\pm0.08) t$ and
$-(0.28\pm0.08)t$, respectively, so that we expect a critical inverse
temperature $\beta t\approx 2.5$\,--\,5,%
\footnote{The relatively large statistical errors of $\Delta E$ and the
  resulting uncertainties in the estimate of the critical temperature are due
  to the addition of the absolute errors to obtain the error of the (small)
  binding energy.}
in agreement with
Figs.~\ref{fig:rho_temp}(a) and (c).
In contrast, the small bipolaron in Fig.~\ref{fig:rho_temp}(b) has a
significantly larger binding energy $\Delta E\approx-(3.43\pm0.09)t$, and
therefore remains stable up to $\beta t\approx 0.3$.

Since the thermal dissociation of intersite bipolarons has been proposed to
explain the activated dc conductivity in the paramagnetic state of the
manganites,\cite{AlBr99prl,AlBr99} we would like to comment on the relation
of our findings to this theory. Instead of the Holstein-type model used
here, Alexandrov and Bratkovsky\cite{AlBr99prl,AlBr99} argue in favor of a model with long-range
el-ph interaction, and assume that charge carriers are O p holes rather than
Mn d holes, so that the double-exchange mechanism does not come into
play. %
\footnote{A discussion of the validity of the two pictures has been given by
  Edwards.\cite{David_AiP}}
An intersite bipolaron in their theory is formed by
two holes residing on neighboring oxygen ions. Furthermore, they also include
a nearest-neighbor Coulomb repulsion $V\approx\Ep$ between
electrons. In the present case, the latter would, most importantly, reduce the binding
energy of the intersite state, thereby leading to a lower temperature for
dissociation. For sufficiently large $V$, intersite bipolaron formation is
expected to be completely suppressed.  A closer investigation of this issue
in the framework of the Holstein-Hubbard model may be carried out using a
generalization of the present method.

In total, a quantitative comparison of our numerical results to the work of
Refs.~\onlinecite{AlBr99prl,AlBr99}, and to the 3D manganites, appears to be
not justified due to the simplifications made and the different model studied
in the present work.

%
%
\subsection{Variational approach}\label{sec:res_vpa}
%
%

%
\begin{figure}
  \includegraphics[width=0.45\textwidth]{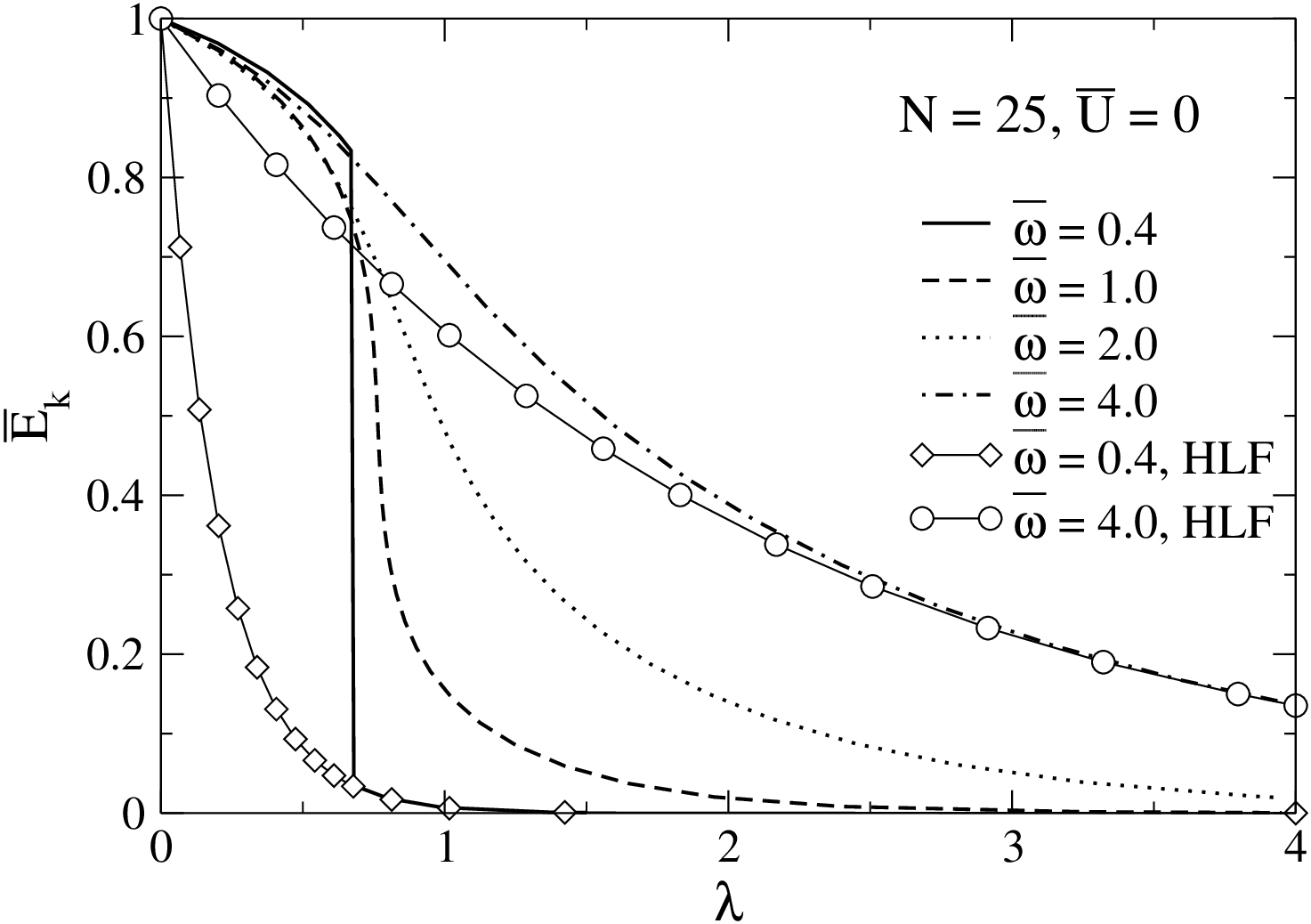}
 \caption{\label{fig:vpa_Ek}
   Variational results for the normalized kinetic energy $\Ekb$ as a function
   of the el-ph coupling $\lambda$, and for different phonon
   frequencies $\omb$. Also shown are results of the HLF approximation (see text).}
\end{figure}

While the above QMC approach is limited to finite temperatures and relatively
small clusters, the variational method of Sec.~\ref{sec:VPA} yields ground-state
results on much larger systems. It becomes exact in several limits. First, for
$\lambda=0$ (\ie, no el-ph coupling), we obtain the exact solution
$\gamma_{ij}=0$ for all $i$, $j$. Second, as $\om\rightarrow\infty$, no
phonons can be excited so that the use of a zero-phonon basis is justified.
Similarly, in the classical limit $\om=0$, the phonons do not have any
dynamics, and the variational determination of the displacement fields allows
one to obtain exact results for any $\lambda$. In contrast, the HLF
approximation (see Sec.~\ref{sec:VPA}) generally overestimates the 
displacement of the lattice at a given site in the presence of an electron,
even for $\om=\infty$. Finally, the variational approach becomes exact in the
nonadiabatic strong-coupling limit $\lambda,\om\rightarrow\infty$. Since the
two-electron problem is diagonalized exactly without phonons, the above
statements hold for any value of the Hubbard repulsion $U$.

\begin{figure}
  \centering
  \includegraphics[width=0.45\textwidth]{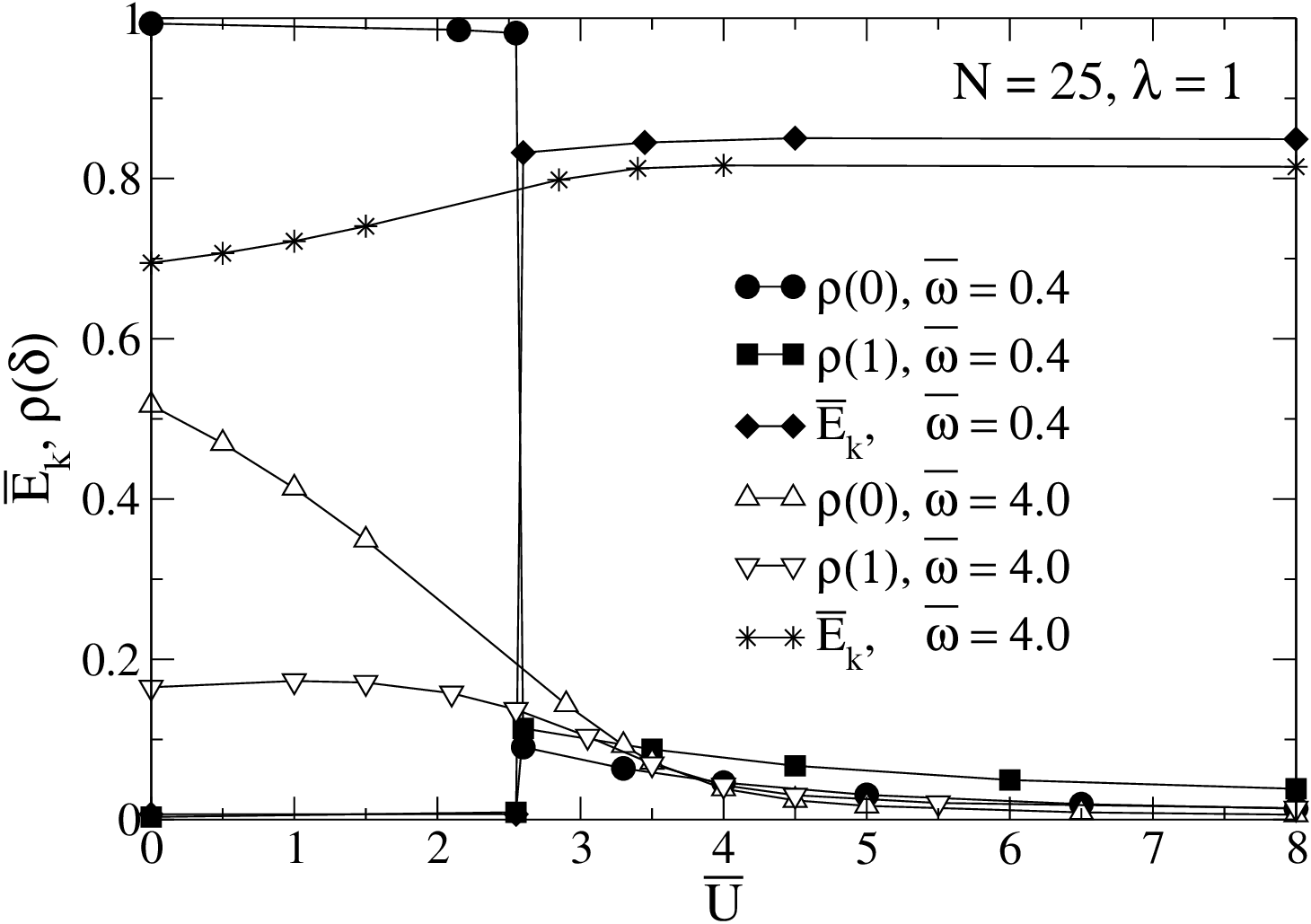}
 \caption{\label{fig:vpa_Ek_rho_U}
   Variational results for the normalized kinetic energy $\Ekb$ and the
   correlation functions $\rho(0)$, $\rho(1)$ as a function of the on-site repulsion $\Ub$.}
\end{figure}

To scrutinize the quality of the variational method, we started by
comparing the ground-state energy for $\Ub=0$ as a function of el-ph coupling for
different values of $\omb$, to the most accurate approach currently available
in one dimension, namely the variational diagonalization.\cite{BoKaTr00} We
find a good agreement over the whole range of $\lambda$. As expected from the
nature of the approximation, slight deviations occur for $\omb\lesssim1$.

Despite the success in calculating the total energy---being the quantity that
is optimized---one has to be careful not to overestimate the validity of any
variational method.  To reveal the shortcomings of the current approach, we show in
Fig.~\ref{fig:vpa_Ek} the normalized kinetic energy $\Ekb=t_\text{eff}$ [see
Eqs.~(\ref{eq:t_eff}) and~(\ref{eq:ekeff})] as
a function of el-ph coupling, and for different $\omb$. We have
chosen $N=25$ to ensure negligible finite-size effects. In principle,
Fig.~\ref{fig:vpa_Ek} displays a behavior similar to the QMC data in
Fig.~\ref{fig:Ek_lambda_omega}(a). There is a strong reduction of $\Ekb$ near
$\lambda=0.5$ for $\omb=0.4$, which becomes washed out and moves to larger
$\lambda$ with increasing phonon frequency. Compared to the exact QMC results
in Fig.~\ref{fig:Ek_lambda_omega}(a), the crossover to a small bipolaron is
much too steep in the adiabatic regime, regardless of the fact that the
variational results are for $T=0$. This is a common defect of variational
methods. Moreover, for $\omb=0.4$\,--\,2.0, the
variational kinetic energy is too small above the bipolaron crossover
compared to the QMC data, while for $\omb=4$, the decay of $\Ekb$ with
increasing $\lambda$ is too slow. 

The reason for the failure is the absence
of retardation effects, which play a dominant role in the formation of
bipolaron states. The increased importance of the phonon dynamics---not
included in the variational method---for the two-electron problem leads to a
less good agreement with exact results than in the one-electron
case.\cite{HoEvvdL03} In particular, our variational results overestimate the
position of the crossover (Fig.~\ref{fig:vpa_Ek}) compared to the
value $\lambda_\text{c}=0.5$ expected in the adiabatic regime.  Nevertheless,
the method represents a significant improvement over the simple HLF
approximation, due to the variational determination of the parameters
$\gamma_{ij}$. This is illustrated in Fig.~\ref{fig:vpa_Ek}, where we also
show the HLF result $\Ekb = \exp(-\Ep/\om)$ for $\omb=0.4$ and 4.0. In contrast to the
variational approach, the HLF approximation yields an exponentially decaying
kinetic energy for all values of the phonon frequency.  Whereas such behavior
actually occurs in the nonadiabatic limit $\om\rightarrow\infty$, the
situation is different for small $\omb$ [see Figs.~\ref{fig:Ek_lambda_omega}(a)
and~\ref{fig:vpa_Ek}]. Thus the HLF approach cannot reproduce the physics of
bipolaron formation for small and intermediate phonon frequencies, while the
variational method presented here accounts qualitatively for the dependence
on the phonon frequency.

Next, we wish to study the influence of Coulomb repulsion $\Ub$. Similar to
Fig.~\ref{fig:S0S1}, we take $\lambda=1$, so that an on-site bipolaron state
is formed at $\Ub=0$. For small phonon frequency $\omb=0.4$,
Fig.~\ref{fig:vpa_Ek_rho_U} reveals a sharp crossover near $\Ub=2.5$, \ie,
at a smaller value of $\Ub$ than in the QMC results of
Fig.~\ref{fig:S0S1}, the reason again being the neglect of the retarded
nature of the effective el-el interaction. As in the QMC results, the Coulomb
repulsion breaks up the on-site bipolaron, leading to an increase of the
kinetic energy. Moreover, the curve for $\rho(1)$
peaks at the crossover point, indicating the existence of an intersite
bipolaron in this regime. A similar picture is found for larger phonon
frequency $\omb=4$, also shown in Fig.~\ref{fig:vpa_Ek_rho_U}, although the
changes with increasing $\Ub$ are much more gradual than for $\omb=0.4$.

\begin{figure}
  \centering
  \includegraphics[width=0.45\textwidth]{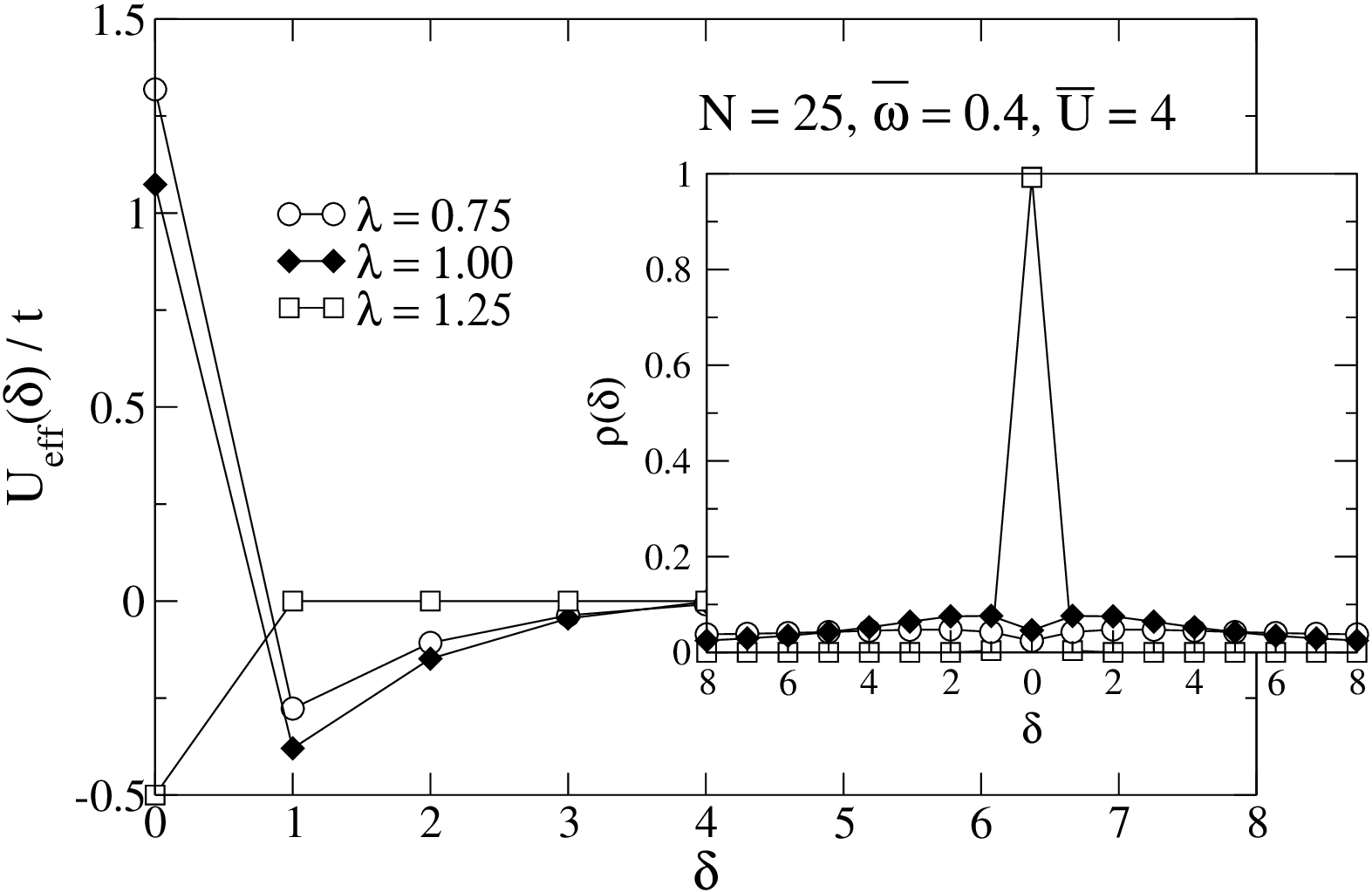}\\
  \includegraphics[width=0.40\textwidth]{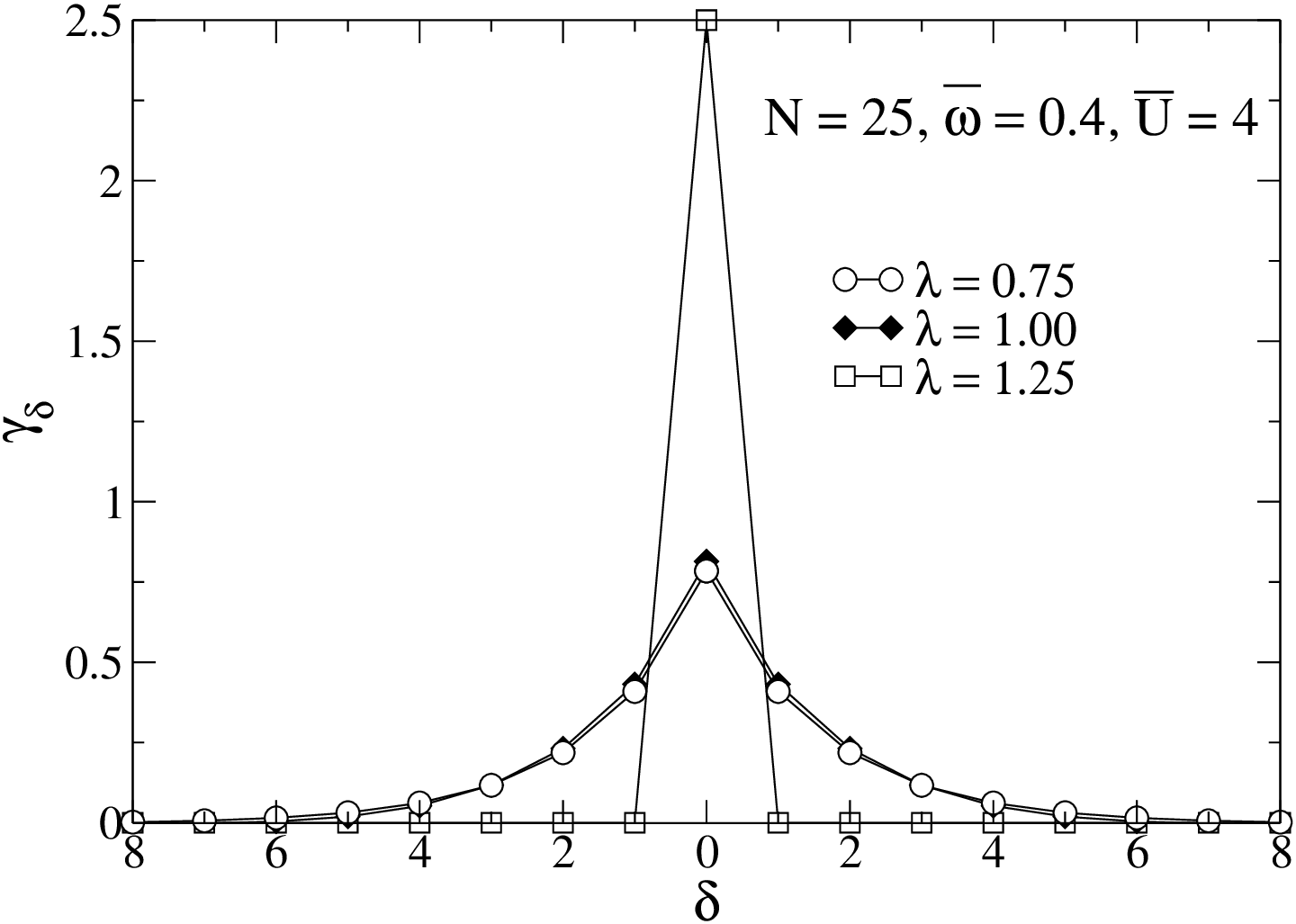}
 \caption{\label{fig:vpa_rho_Ueff}
   Upper panel: variational results for the effective interaction
   $U_\text{eff}(\delta)$ (see text) and the correlation function
   $\rho(\delta)$ (inset) as a function of the el-el distance $\delta$.
   Lower panel: variational lattice distortions $\gamma_\delta$ as a function of
   $\delta$.
}
\end{figure}

Finally, we report in Fig.~\ref{fig:vpa_rho_Ueff} (upper panel) the effective
interaction $U_\text{eff}(\delta)$ between the two electrons as a function of
their relative distance $\delta$, given by $v_\delta$
[Eq.~(\ref{eq:htilde_el-el})]. We have chosen $\omb=0.4$ and
$\Ub=4$, the same parameters as in Fig.~\ref{fig:rho}(b). For
$\lambda=0.75$, the finite Coulomb repulsion stabilizes two weakly bound
polarons, as illustrated by the results for $\rho(\delta)$ shown in the inset of
Fig.~\ref{fig:vpa_rho_Ueff}. While $U_\text{eff}$ is repulsive
(positive) for $\delta=0$, the two electrons can form a bound state by
traveling a finite distance $1\leq\delta\lesssim4$ apart. This is still true
for $\lambda=1$, for which the HLF approach yields
$U_\text{eff}(0)=U-2\Ep=0$. Nevertheless, the two electrons experience an
attractive interaction and form an extended bipolaron. Finally, for even
stronger coupling $\lambda=1.25$, the phonon-mediated el-el
interaction has overcome the on-site repulsion, so that
$U_\text{eff}(\delta=0)<0$. At the same time, the size of the bipolaron has
collapsed to a single site. 
This crossover is also well visible in the lower
panel, which displays the variationally determined lattice distortions
$\gamma_\delta$.
It is worth mentioning that the values of
$U_\text{eff}(0)$ in Fig.~\ref{fig:vpa_rho_Ueff} are larger than
the strong-coupling prediction $U-2\Ep$ for all values of $\lambda$
considered.  This may be attributed to the overestimated bipolaron binding
energy in the atomic limit.

As pointed out in several places, the shortcomings of the variational
approach presented here are a result of the missing dynamical phonon
effects. The present approach may be further improved by
making an ansatz for the eigenstates of the untransformed
Hamiltonian~(\ref{eq:holstein}) of the form
\begin{equation*}
  \begin{split}
  \ket{\Psi_k}
  =
  \frac{1}{N}
  \sum_{ij}
  &
  \sum_p
  e^{\rmi p x_i + (k-p)x_j}
  \\
  &\times
  \left(
  \tilde{d}^{(1)}_p\nu^\dag\{\bm{\gamma}^{(1)}\}
  +
  \tilde{d}^{(2)}_p\nu^\dag\{\bm{\gamma}^{(2)}\}
  \right)
  \ket{i,j}
   \,, 
  \end{split}
\end{equation*}
with $\ket{i,j}$ defined as in Eq.~(\ref{eq:basis}), two canonical
transformations depending on the displacement fields $\gamma^{(1)}_{ij}$ and
$\gamma^{(2)}_{ij}$ (see Sec.~\ref{sec:LFT}), and additional variational
parameters $\tilde{d}^{(1)}_p$, $\tilde{d}^{(2)}_p$. Thereby, one can take into
account lattice distortions not centered at the sites of the electrons, which
become important as $\omb\rightarrow0$, and which reproduce to some degree
the effect of retardation.

%
%
%
%
\section{\label{sec:summary}Conclusions}
%
%
%

We have studied the Holstein-Hubbard bipolaron with quantum
phonons by extending a quantum Monte Carlo method previously developed for
the Holstein polaron.\cite{HoEvvdL03} In its present form, the method is
limited to one-dimensional clusters. However, in contrast to other
approaches, it allows to perform accurate calculations also
for small phonon frequencies and finite temperatures.

We have studied the dependence of bipolaron formation on the phonon frequency
and the Hubbard repulsion. Our results underline the importance of the phonon
dynamics, which has often been neglected in previous work. Moreover, we have
presented for exact results for the effect of temperature on
the bipolaron state in the important adiabatic regime. 
Thermal dissociation of bipolarons has been observed at temperatures
where the thermal energy becomes comparable to the binding energy.

Two interesting open issues are the effect of nearest-neighbor Coulomb
interaction, as well as that of dimensionality. While the latter cannot
easily be addressed with the current approach, one may instead extend the
promising work of Ref.~\onlinecite{deRaLa86} to finite phonon
frequencies.

Finally, we have proposed a variational ansatz based on a canonical
transformation with variational parameters. The latter represents a
significant improvement over standard approximations. In particular, it
qualitatively accounts for the dependence on the phonon frequency.

%
%
%
\begin{acknowledgments}
%
%
%
  
  This work has been supported by the Austrian Science Fund (FWF), project
  No.~P15834.  One of us (M.~H.) is indebted to DOC (Doctoral Scholarship
  Program of the Austrian Academy of Sciences). We would like to thank Holger
  Fehske for useful discussion and valuable comments on the manuscript.
  
%
%
%
\end{acknowledgments}
%
%
%




\end{document}